\documentclass[12pt]{iopart}

\usepackage{graphicx}
\usepackage{caption}
\usepackage{subcaption}
\usepackage{braket}
\usepackage{cite}
\usepackage{longtable}
\usepackage{amsfonts}
\usepackage[dvipsnames]{xcolor}
\usepackage{sidecap,tikz, multirow,subcaption} 
\usepackage{braket}
\expandafter\let\csname equation*\endcsname\relax
\expandafter\let\csname endequation*\endcsname\relax
\usepackage{amsmath}
\usepackage{array}
\usepackage{xcolor}
\usepackage[breaklinks=true,colorlinks=true,linkcolor=blue,urlcolor=blue,citecolor=blue]{hyperref}
\usepackage{bm,dsfont}
\usepackage[sort&compress,numbers]{natbib}
\newcommand{\newblock}{}
\begin{document}

\title{Fluid Simulation for a 
 Finite Size Plasma}

\author{Subhasish Bag}
\address{Department of Physics, Indian Institute of Technology Delhi, Hauz Khas, New Delhi 110016, India}
\ead{subhasishbag95@gmail.com}

\author{Vikrant Saxena}
\address{Department of Physics, Indian Institute of Technology Delhi, Hauz Khas, New Delhi 110016, India}
\ead{vsaxena@physics.iitd.ac.in}

\author{Amita Das}
\address{Department of Physics, Indian Institute of Technology Delhi, Hauz Khas, New Delhi 110016, India}
\ead{amita@iitd.ac.in}

\begin{abstract}
 Studies on finite-size plasma have attracted a lot of attention lately. They can form by ionizing liquid droplets by lasers. The dynamical behavior of such plasma droplets is, therefore,  a topic of significant interest. In particular, questions related to the linear and nonlinear characteristics (associated with the inhomogeneous density typically at the edge of the droplet), the behavior of plasma expansion, etc., are of interest. A one-dimensional fluid simulation study has been carried out to investigate this behavior. It is observed that a slight imbalance in the charge density leads to oscillations that are concentrated and keep acquiring higher amplitude and sharper profile at the inhomogeneous edge region. Such oscillations lead to the expansion of the droplet. Though the fluid description breaks when the sharpness of these structures becomes comparable to the grid size, it provides a reasonable estimate of wave-breaking time. The presence of dissipative effects like diffusion is shown to arrest the sharpness of these structures. The dynamics of these structures in the presence of an externally applied oscillating electric field corresponding to a long wavelength radiation has also been studied. 
 
 \end{abstract}

\section{Introduction:}
The nonlinear plasma oscillations and waves have been widely investigated for the past several decades due to their relevance in the context of charged particle acceleration\cite{tajima1979laser, malka2002electron, modena1995electron, faure2006controlled, rechatin2009controlling}, fast ignition\cite{tabak1994ignition}, particle heating\cite{pukhov1996relativistic}, radiation sources\cite{anand2007laser}, etc.   These oscillations have a threshold amplitude ( Dawson\cite{dawson1959nonlinear} and Davidson-Schram\cite{davidson1968nonlinear}) beyond which they undergo wave breaking. Several other studies \cite{albritton1975relation, verma2011nonlinear, sengupta2011phase} have identified the phenomena of phase mixing which also destroys the sanctity of the wave. The phase mixing process happens when the plasma frequency becomes a function of space. This may happen as a result of inhomogeneous plasma density or relativistic effects which alter the mass of the particles. In such a scenario the plasma oscillations lose phase coherence which leads to the breaking of waves.  Thus phase mixing can lead to wave breaking even if the threshold amplitude is not reached \cite{dawson1959nonlinear, kaw1973quasiresonant, infeld1989ion}. The effect of finite ion mass\cite{gupta1999phase, verma2011nonlinear} and relativistic effect \cite{sengupta2009phase, bera2016relativistic, bera2021effect} have been explored in the context of phase mixing phenomena. 
Earlier studies on wave breaking and phase mixing have, however, been carried out for the infinitely extended plasma medium. In this work, we explore the behavior of oscillations when the plasma has a finite size.  For a finite plasma having a density profile that is more or less uniform in the central region but falls steeply but continuously at the edges, the frequency spectrum is partly continuous ( consisting of frequencies corresponding to the continuously falling plasma density in the edge region), and partly discrete (plasma frequency corresponding to the uniform plasma density). Barston et al.\cite{barston1964electrostatic} have shown using a normal mode analysis that for a finite plasma with sharp but continuous edges there must exist a well-behaved mode in the edge region with a frequency $\omega$ satisfying $\omega^2<\omega^2_{p0}$. The condition for a transition region between two different plasma densities becomes $\omega^2_{p2}< \omega^2<\omega^2_{p1}$ where $\omega_{p1}$ and $\omega_{p2}$ are plasma frequencies corresponding to the two plasma densities separated by the transition region.

  We have developed a two-fluid simulation model for cold plasma studies. We investigate the oscillations and also study the expansion of the plasma. The cold plasma limit is valid when the plasma wave intensity satisfies the condition  $|E|^2/4\pi{n}T>>1$. Here $E$ represents the electric field amplitude and $ n$ and $T$ correspond to the density and temperature, respectively, of the plasma. The magnetic field is assumed to be absent and the amplitude of the plasma wave is low for relativistic effects to have any role. Both cases  (sharp and smooth boundaries) of the finite-size plasma have been considered. 

 We observe the development of several sharp structures at the plasma edge in the numerical simulations. These structures keep getting sharper as they evolve and move down the density gradient. 
 The article has been organized as follows. In sec-\ref{2s}, we have discussed the analytical model for our plasma system, along with the simulation details that have been employed. In sec-\ref{3s}, the observations for infinitely extended plasma for benchmarking and testing have been first carried out.  Thereafter, studies on some finite-size plasma profiles have been presented. The formation of several sharp structures at the inhomogeneous edge region of the plasma is evident from the simulation results. The role of diffusion in capturing the sharpness of these structures has been investigated in sec-\ref{4s}. In sec-\ref{5s} the driven plasma system is considered.  
 Finally, we conclude our studies in sec-\ref{6s}.

\section{Fluid Model and Simulation Details:}{\label{2s}}
We consider a two-fluid electrostatic plasma model for cold, collisionless, and unmagnetized plasma. The one-dimensional two-fluid equations of our model are,
\begin{equation} \label{eq:1}
    {\frac{\partial{n_e}}{\partial{t}}}+{\frac{\partial{(n_eu_e)}}{\partial{x}}}=0 
\end{equation}
\begin{equation} \label{eq:2}
    {\frac{\partial{n_i}}{\partial{t}}}+{\frac{\partial{(n_iu_i)}}{\partial{x}}}=0 
\end{equation}
\begin{equation} \label{eq:3}
    {\frac{\partial{u_e}}{\partial{t}}}+u_e{\frac{\partial{u_e}}{\partial{x}}}={\frac{e}{m_e}}\frac{\partial{\phi}}{\partial{x}}
\end{equation}
\begin{equation} \label{eq:4}
    {\frac{\partial{u_i}}{\partial{t}}}+u_i{\frac{\partial{u_i}}{\partial{x}}}=-{\frac{e}{m_i}}\frac{\partial{\phi}}{\partial{x}}
\end{equation}
\begin{equation} \label{eq:5}
    \frac{\partial^2 {\phi}}{\partial {x}^2}=\frac{e}{\epsilon_0}({n_e}-{n_i})
\end{equation}
The symbols $n$ and $u$ are the fluid density and fluid velocity. The subscripts $e$ and $i$ represent the electron and ion fluids, respectively. The electrostatic potential has been depicted by $\phi$,  $e$ is the magnitude of the electronic charge, $\epsilon_0$ is the vacuum permittivity, and the ion and electron masses have been represented by $m_i$ and $m_e$, respectively. Equations (\ref{eq:1}) and (\ref{eq:2}) are the electron and ion fluid continuity equations, respectively, while equations (\ref{eq:3}) and (\ref{eq:4})  express the electron and ion fluid momentum equations, respectively. Equation (\ref{eq:5}) is the one-dimensional Poisson equation.

A one-dimensional fluid simulation has been carried out to solve the set of fluid equations using the flux corrected scheme (FCT)\cite{boris1993lcpfct, boris1976solution}. Along with these equations, a 1D Poisson solver (cyclic tridiagonal method\cite{press1992numerical}) has been used to solve the Poisson equation which couples the electron and ion fluids. We have normalized our variables in equations (\ref{eq:1}-\ref{eq:5}) as shown  in table (\ref{tab1}). 
\begin{table}[h]
    \centering
$$\begin{tabular}{|c|c|c|}
\hline
Variables  &{Normalized by}  &{Normalized form}    \\ \hline
Fluid Velocity ({$u$}) & Maximum Electron Velocity ($v_{max}$) &$\Bar{u}_e,\Bar{u}_i$  \\
Number Density (n) &{ Maximum Density ($n_0$)} &$\Bar{n}$  \\ 
Time (t) &{$\omega_{pe}^{-1}$ where, $\omega_{pe}=\sqrt{n_0e^2/{m_e\epsilon_0}}$} &$\Bar{t}$  \\ 
Position ({x}) &{($v_{max}/{\omega_{pe}}$)} &$\Bar{x}$  \\ 
Potential ($\phi$) &${m_ev_{max}^2}/e$ &$\Bar{\phi}$  \\ \hline
\end{tabular}$$\\
    \caption{Normalized variables}
    \label{tab1}
\end{table}
The bar on the normalized variables will be removed henceforth.
The modified equations in normalized variables are 
\begin{equation} \label{eq:6}
    {\frac{\partial{n}_e}{\partial{t}}}+{\frac{\partial{({n}_e{u}_e)}}{\partial{{x}}}}=0
\end{equation}
\begin{equation} \label{eq:7}
    {\frac{\partial{n}_i}{\partial{t}}}+{\frac{\partial{({n}_i{u}_i)}}{\partial{{x}}}}=0
\end{equation}
\begin{equation} \label{eq:8}
    {\frac{\partial{u}_e}{\partial{t}}}+{u}_e{\frac{\partial{u}_e}{\partial{x}}}={\frac{\partial{\phi}}{\partial{x}}}
\end{equation}
\begin{equation} \label{eq:9}
    {\frac{\partial{u}_i}{\partial{t}}}+{u}_i{\frac{\partial{u}_i}{\partial{x}}}=-\frac{m_e}{m_i}{\frac{\partial{\phi}}{\partial{x}}}
\end{equation}
\begin{equation} \label{eq:10}
    \frac{\partial^2 {\phi}}{\partial {x}^2}=({n}_e-{n}_i)
\end{equation}
The spatial grid size and the time step have been chosen to satisfy the Courant condition. 
The simulation box size is about $100$
units and the corresponding  grid size (d${x}$) is 0.01 and the time step (d${t}$) is 0.0001. 

\section{Numerical Observations:}{\label{3s}}
We have first carried out some benchmark studies to ascertain the correctness and accuracy of our simulation. For this purpose known results of homogeneous infinitely extended 1D plasma, represented by a finite box of size $L$ using periodic boundary conditions, have been recovered. 
The initial velocities of both ion and electron are chosen to be zero. The ion density is chosen to be $\bar{n}_i = n_0 = 1$ and is a constant initially.  
A sinusoidal perturbation in electron density is imposed   $\bar{n}_e = 1 + \tilde{n} \sin{k_0 x}$. Here $\tilde {n} <<0.5$ and $k_0 = \frac{2\pi}{L}$. The initial density imbalance between ions and electrons gives rise to an electrostatic field which drives the motion of the charged particles leading to  plasma oscillations occurring at a frequency of  
\begin{equation} \label{eq:11}
    {\omega^2=\omega_{pe}^2+\omega_{pi}^2}
\end{equation}
Here $\omega_{pe} $ and $\omega_{pi}$ are the electron and ion plasma frequencies, respectively. Since the ions are much heavier in comparison to electrons the oscillation frequency was found to be close to the electron plasma frequency. 

The dynamics described by the set of (\ref{eq:6}-\ref{eq:10}) ensure that the total energy comprising of the sum of kinetic and potential energy remains conserved to have   
\begin{equation} \label{eq:12}
    \frac{d}{d{t}}\int\left[\frac{1}{2}{n}_e{u}_e^2+\frac{1}{2}\frac{m_i}{m_e}{n}_i{u}_i^2+\frac{1}{2}\left(\frac{\partial{\phi}} {\partial{x}}\right)^2\right]d{x}= \frac{dE_t}{dt} = 0
\end{equation}
Here $E_t$ is the total energy.  We have verified the conservation of total energy for our program. The conservation of total energy is an extremely important tool and is utilized for checking the validity and accuracy of the simulations that are being carried out. 
\begin{figure}[!hbt]
     \begin{subfigure}[b]{0.5\textwidth}
         \includegraphics[scale=0.4]{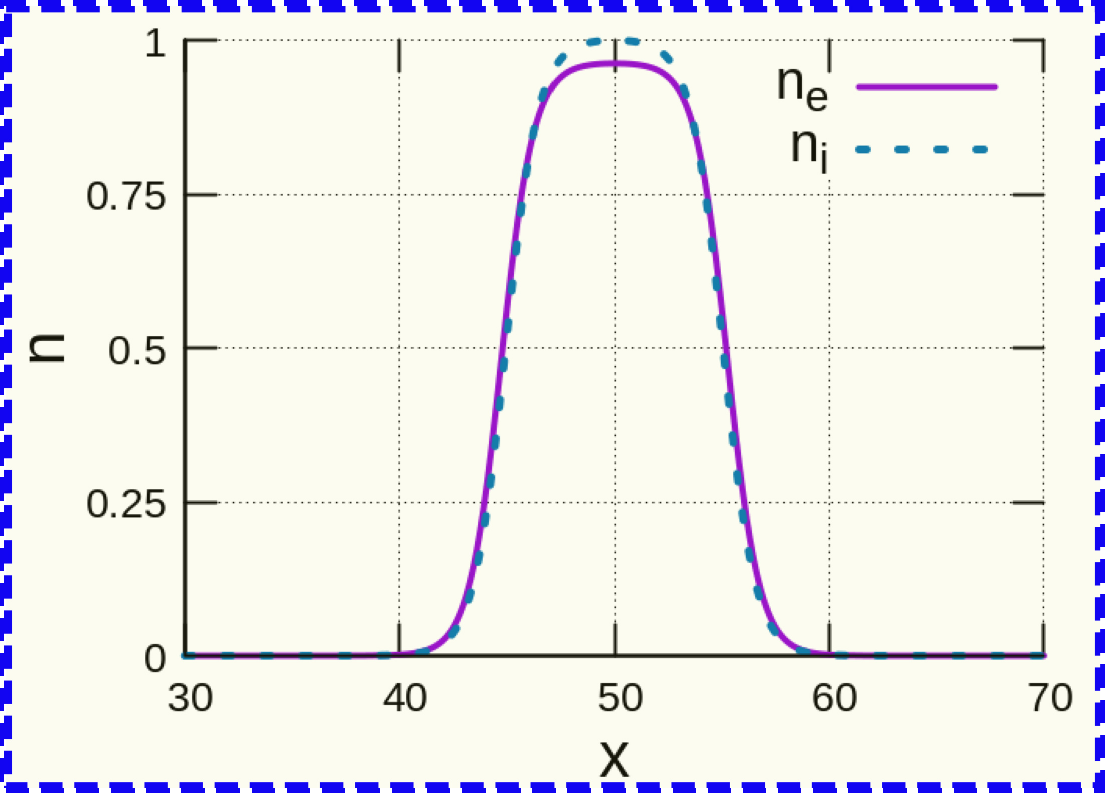}
         \caption{}
         \label{fig:1a}
     \end{subfigure}
        \begin{subfigure}[b]{0.5\textwidth}
         \includegraphics[scale=0.4]{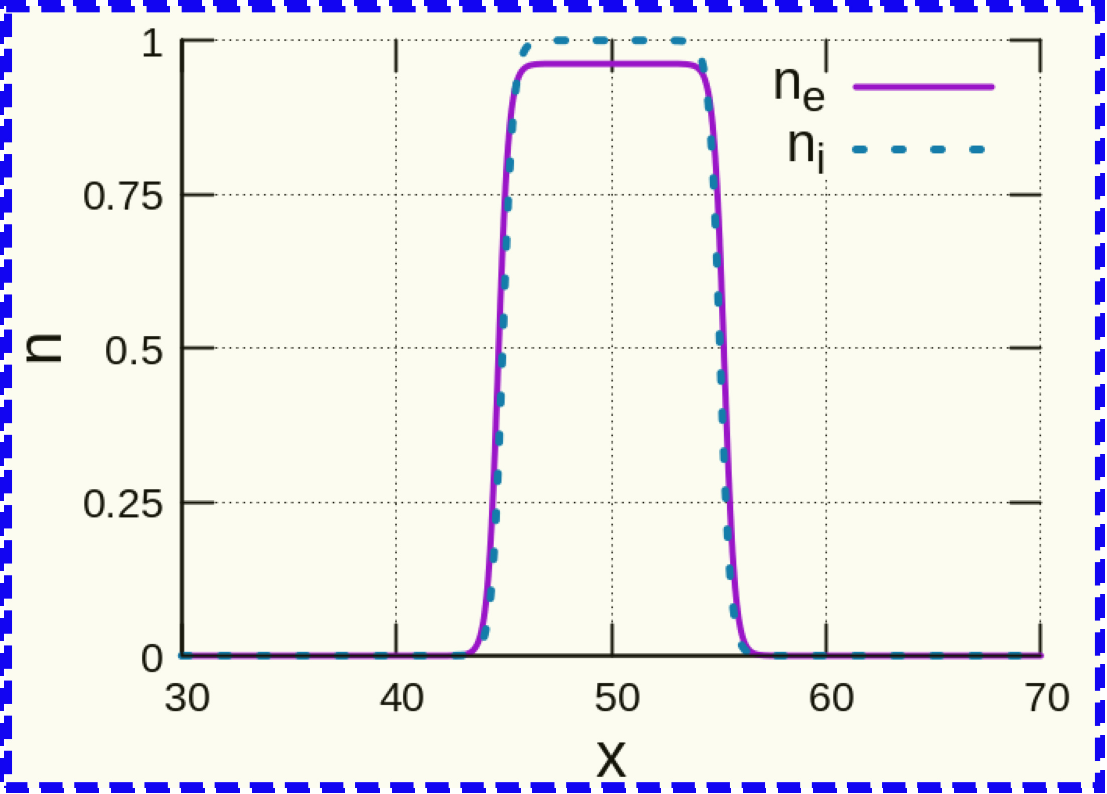}
         \caption{}
         \label{fig:1b}
     \end{subfigure}
        \caption{Initial Density profiles. (\ref{fig:1a}) has the smoother edge and (\ref{fig:1b}) sharper edge.}
        \label{fig:1}
\end{figure}

We now consider a plasma of finite spatial extent. The electron and ion density profiles have been shown in figure ({\ref{fig:1}}). It can be observed that while figure ({\ref{fig:1a}}) has a smooth edge, figure ({\ref{fig:1b}}) has a very sharp edge. These profiles have been constructed using the double tangent hyperbolic functional form given by the equation below 
\begin{equation}\label{eq:13}
    {n}_i(x)=tanh\left(\frac{x-a_i}{\alpha}\right)-tanh\left(\frac{x-b_i}{\alpha}\right) 
\end{equation}
\begin{equation}\label{eq:14}
    {n}_e(x)= (1 - \tilde{n})\times\left[tanh\left(\frac{x-a_e}{\alpha}\right)-tanh\left(\frac{x-b_e}{\alpha}\right)\right]
\end{equation}
The parameter $\alpha$ defines the sharpness of the edge region. Increasing $\alpha$ makes the plasma edge smoother. For the profiles of figure ({\ref{fig:1a}}) and figure (\ref{fig:1b}) the choice for $\alpha $ is $1.5$ and $0.5$ respectively. The other parameters $a_s$, and $b_s$ for $s = i,e$ define the center and the width of the density profile for the respective densities. We have considered the values $a_i=44.95$, $b_i=55.05$, $a_e=44.75$, and $b_e=55.25$ for both cases.  For ions, the value of the maximum density is unity. For electrons, the amplitude and the width are chosen to be slightly different to incorporate a charge imbalance in the initial condition, however, it is ensured that the integrated densities of both species remain the same, to satisfy quasi-neutrality. It can be observed from the figure (\ref{fig:1}) that the ion density in the central region is higher, but its width is smaller compared to the electron density such that the area under the curve is the same.  Furthermore, the double tangent hyperbolic profile also ensures the flat top homogeneous central density.

\begin{figure}[!hbt]
    \centering
    \includegraphics[scale=0.75]{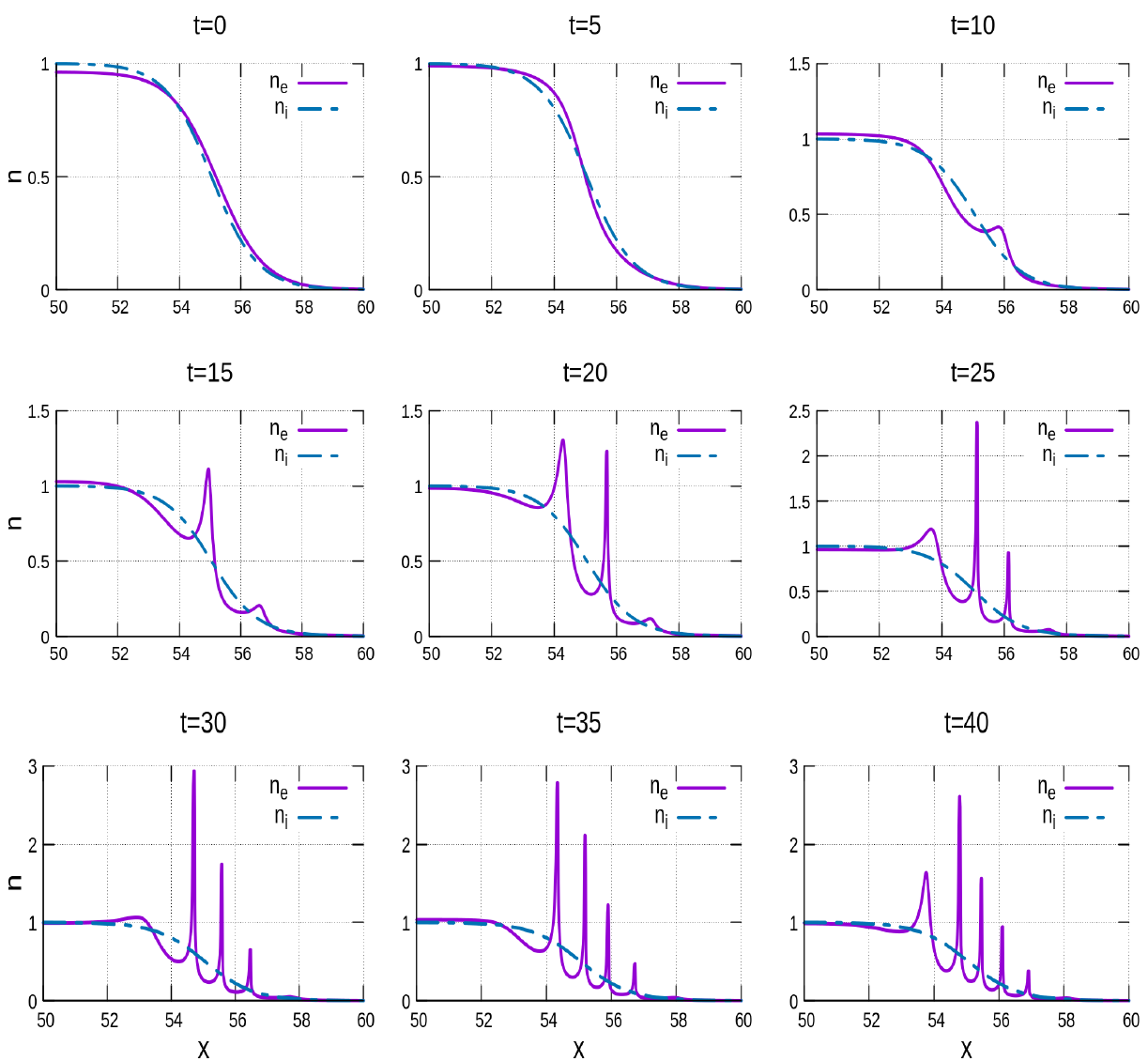}
    \caption{Density Evolution of the case: figure-(\ref{fig:1a}).}
    \label{fig:2}
\end{figure}

\begin{figure}[!hbt]
    \centering
    \includegraphics[scale=0.75]{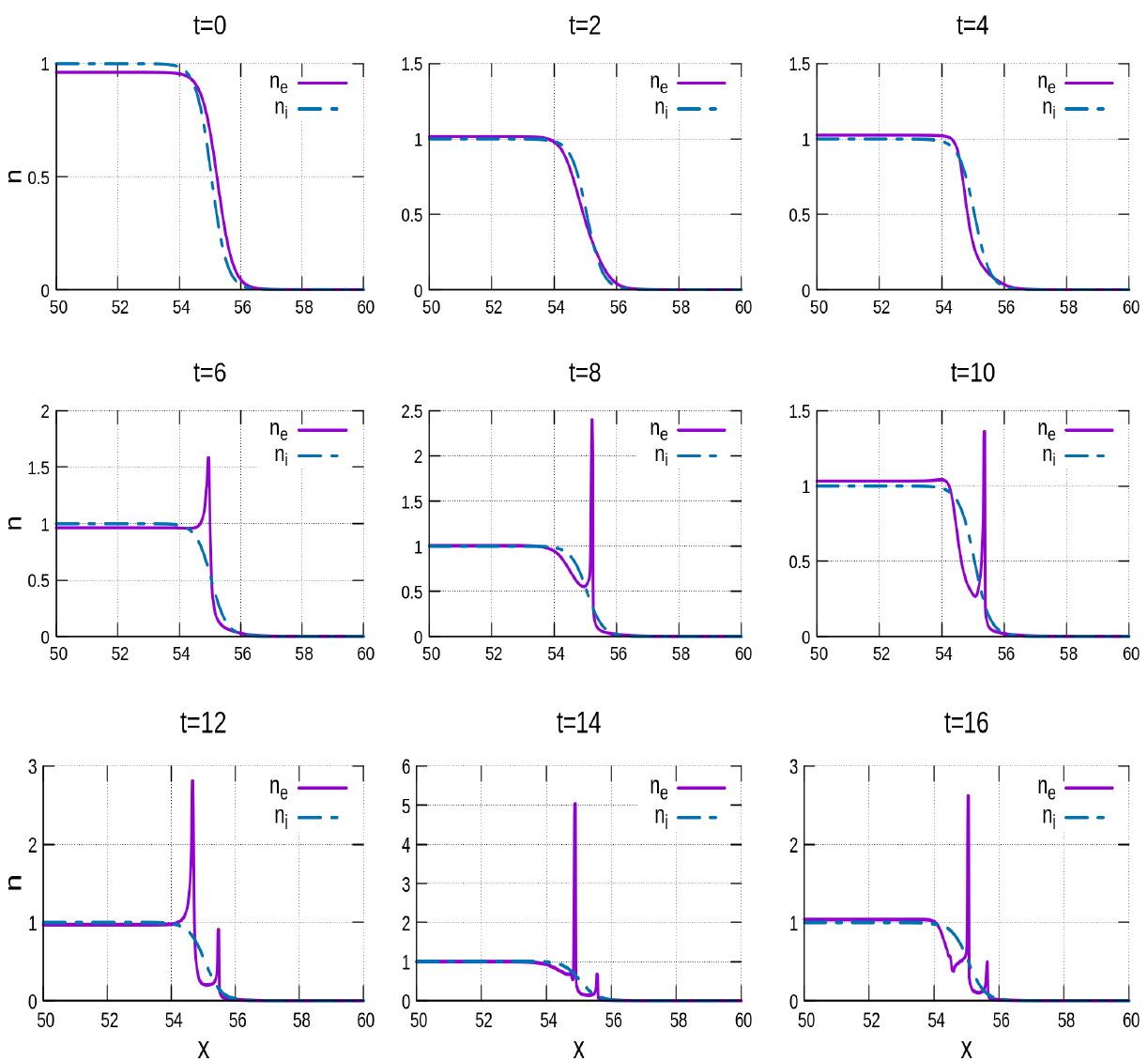}
    \caption{Density Evolution of the case: figure-(\ref{fig:1b}).}
    \label{fig:3}
\end{figure}

\begin{figure}[!hbt]
    \centering
    \includegraphics[scale=0.75]{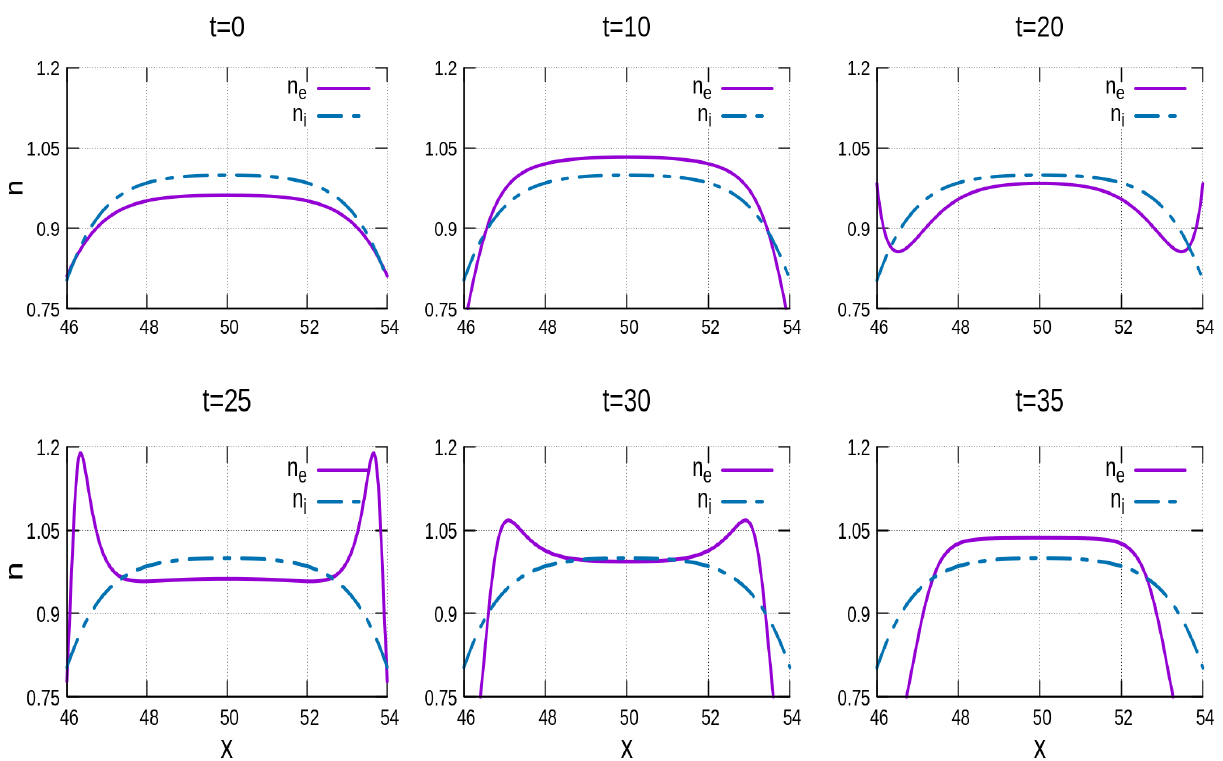}
    \caption{Density Evolution at the central region of the case: figure-(\ref{fig:1a}).}
    \label{fig:4}
\end{figure}

\begin{figure}[!hbt]
    \centering
    \includegraphics[scale=0.75]{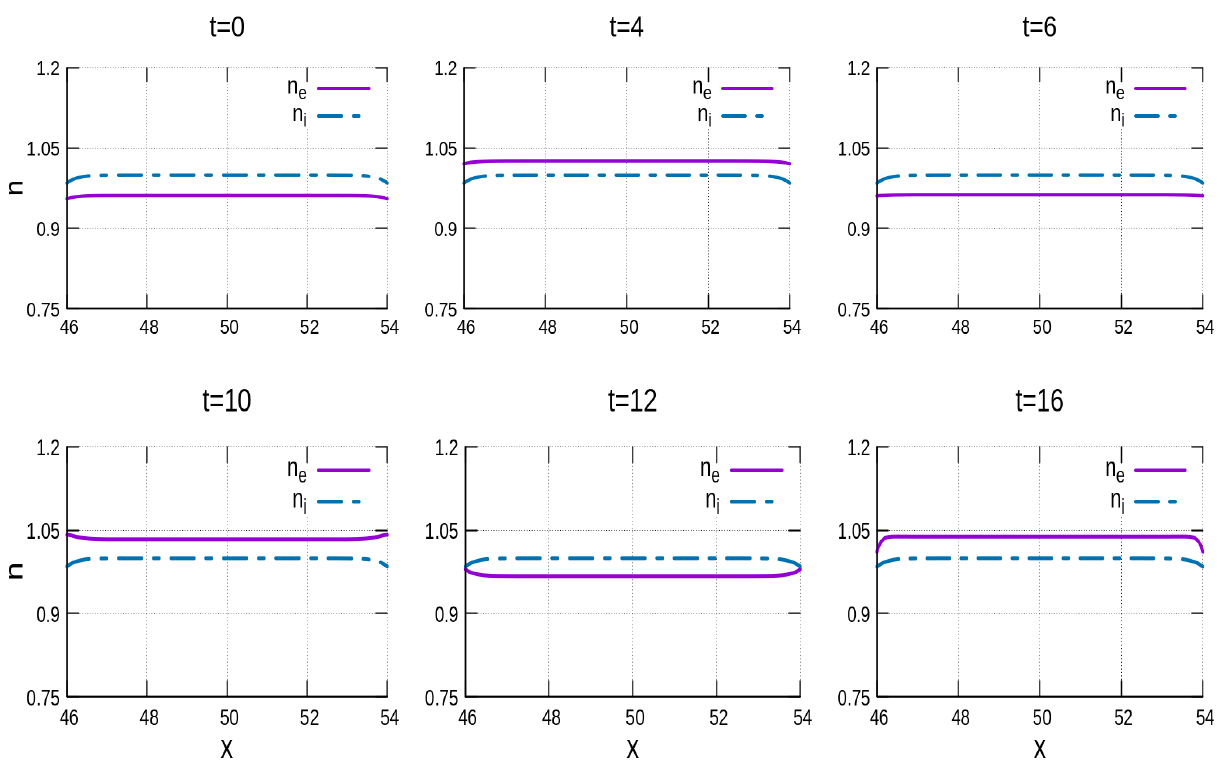}
    \caption{Density Evolution at the central region of the case: figure-(\ref{fig:1b}).}
    \label{fig:5}
\end{figure}

Figure (\ref{fig:2}) shows the time evolution of electron and ion densities for the initial shallow profile depicted in figure (\ref{fig:1a}), at various times.  It is observed that peaks keep appearing in the electron density at the edge of the profile. They subsequently move down the slope and keep getting sharper.  This happens all within a few electron plasma periods. For the sharper background plasma profile of figure (\ref{fig:3}), the peaks that appear are much sharper, though the dynamics remain similar. It should also be noted that there are no sharp structure formations in the central region of the flat density profile. Here, merely the amplitude of electron density keeps oscillating as a whole. The oscillating electron and ion density amplitude at the center has been plotted as a function of time in figures (\ref{fig:4}) and (\ref{fig:5}). The frequency of these oscillations corresponds to the plasma frequency corresponding to the maximum density.
\begin{figure}[!hbt]
     \begin{subfigure}[b]{0.5\textwidth}
         \includegraphics[scale=0.285]{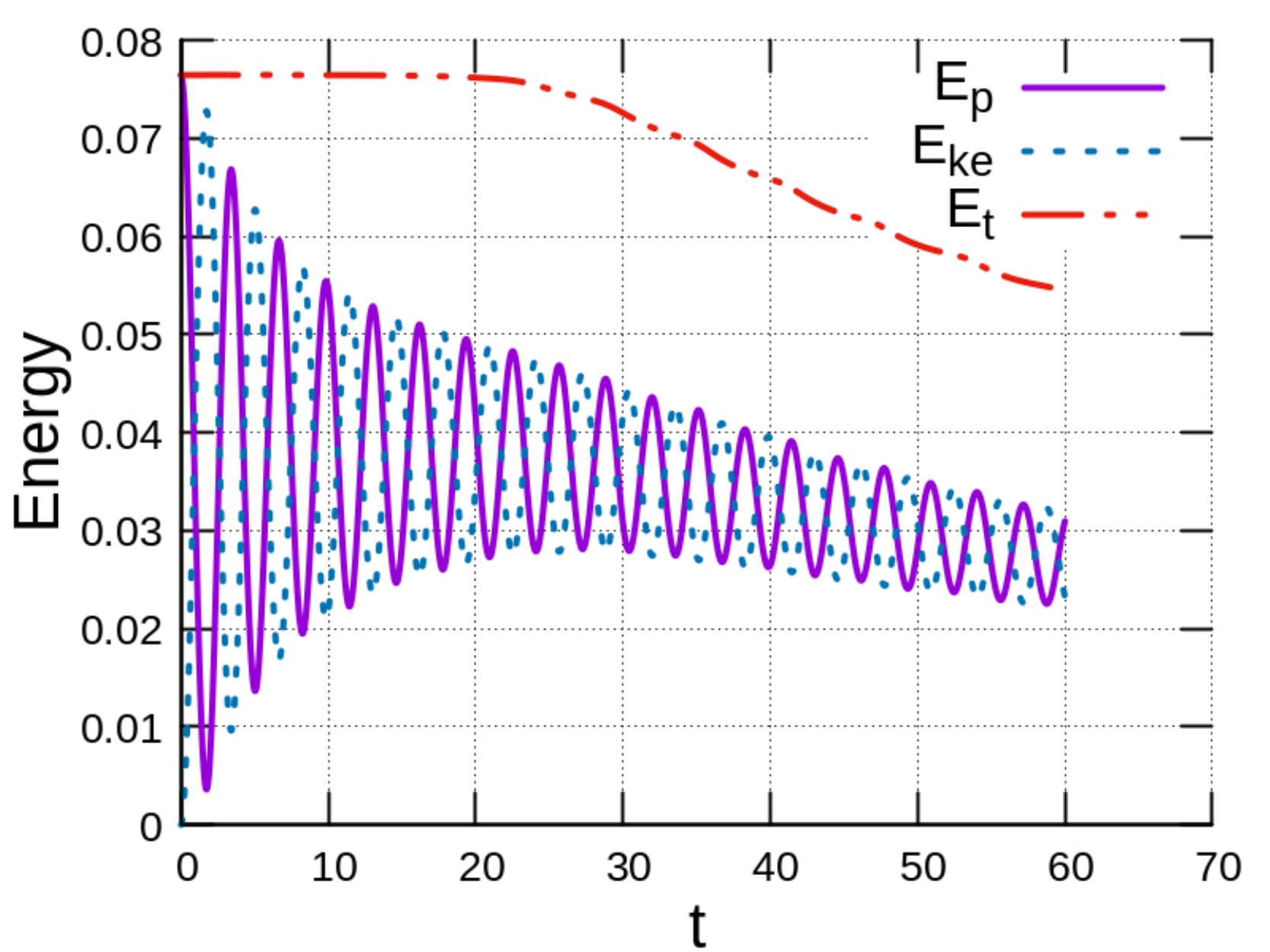}
         \caption{}
         \label{}
     \end{subfigure}
        \begin{subfigure}[b]{0.5\textwidth}
         \includegraphics[scale=0.285]{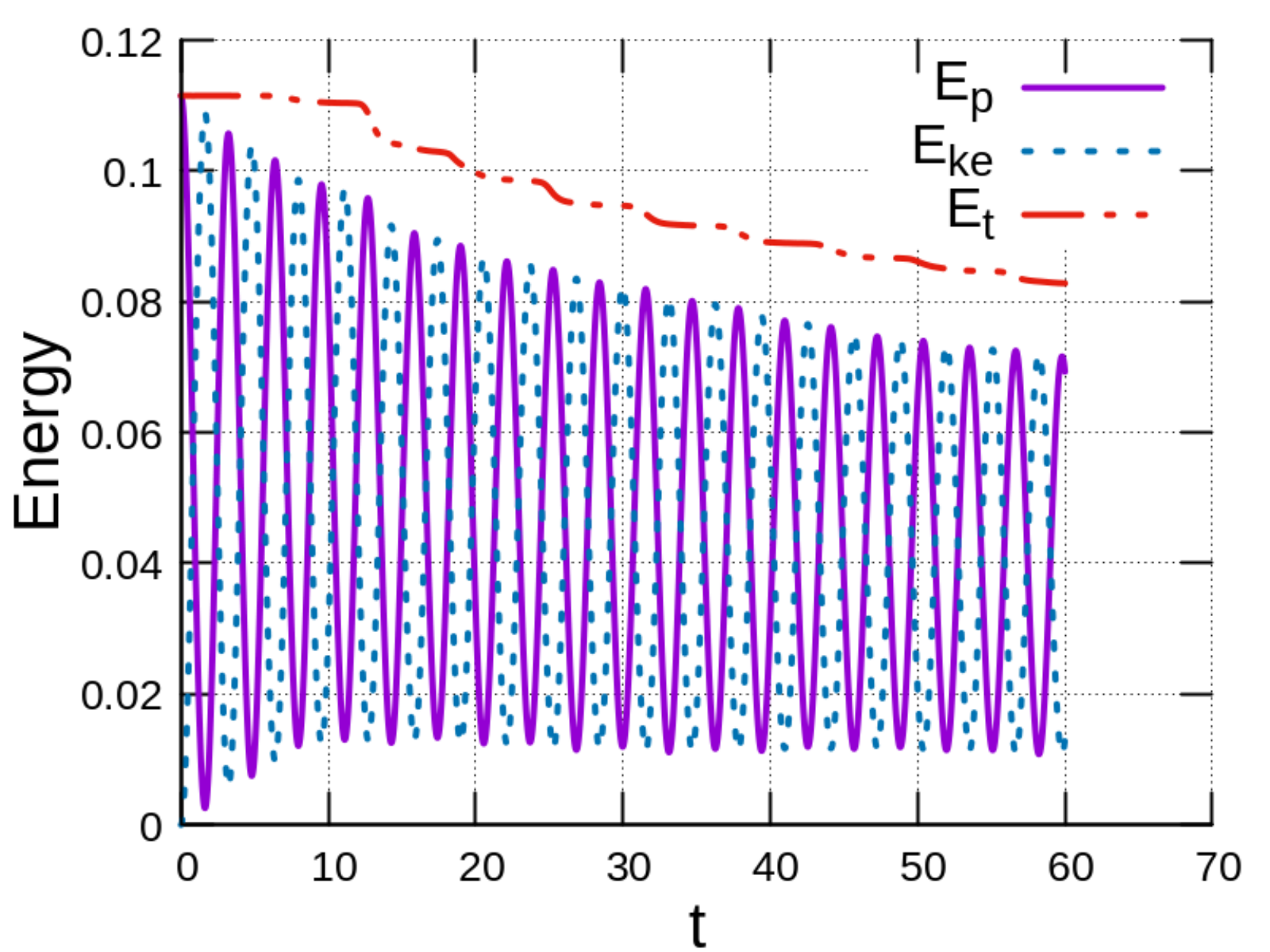}
         \caption{}
         \label{}
     \end{subfigure}
        \caption{Total Energy for two cases (figures-\ref{fig:1a} and \ref{fig:1b}) where $E_p$, $E_{ke}$, $E_{t}$ are the Potential Energy, Electron Kinetic Energy, and Total Energy, respectively.}
        \label{fig:6}
\end{figure}

\begin{figure}
    \centering
    \includegraphics[scale=0.75]{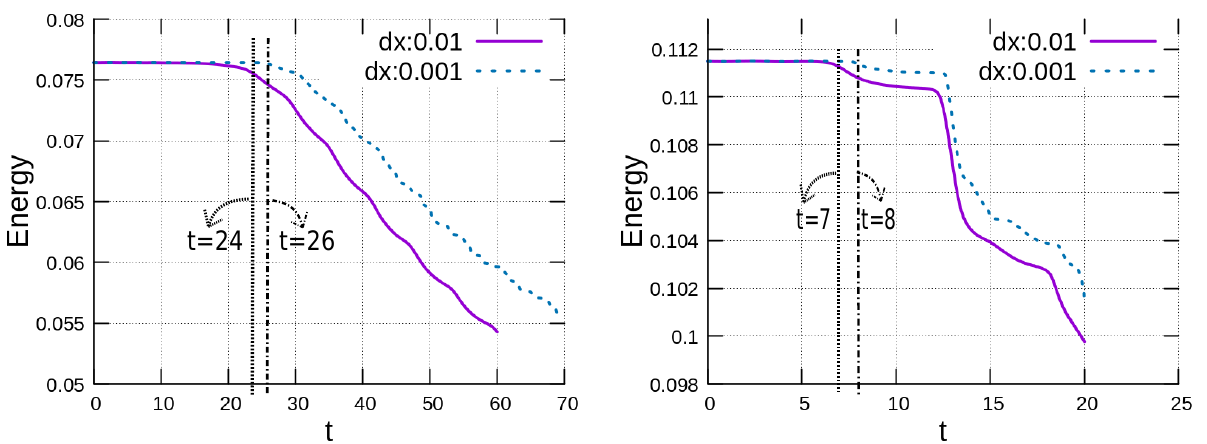}
    \caption{Total energy with different dx for the smooth density case: figure-(\ref{fig:1a}) and the sharp density case: figure-(\ref{fig:1b}).}
    \label{fig:7}
\end{figure}

The plots in figure (\ref{fig:2}) and (\ref{fig:3}) show that the peaks at the edge continue to get sharper. Thus the energy associated with these structures ultimately lands up at the grid scales in fluid simulations. This is the time when the fluid simulations will get invalidated. To discern the duration of applicability of fluid simulations, we follow the evolution of total energy. For the two simulations, the total energy evolution along with the evolution of electron kinetic and the electrostatic potential energy has been shown in the plots of the figure (\ref{fig:6}). The figure clearly shows that the total energy remains conserved only for some initial duration. This time is shorter when the chosen background plasma profile is sharper, for which sharper perturbations appear right from the very beginning. To show that this is indeed related to the finite choice of grid size we have also carried out simulations with a better grid resolution of $dx = 0.001$, a value which is about $10$ times smaller than our previous choice of $dx = 0.01$. A comparison of energy evolution for these two grid spacings has been made in figure (\ref{fig:7})  for both profiles. 
It is evident from the figures that by reducing the grid size, energy conservation remains valid for a slightly longer duration.

\begin{figure}
    \centering
    \includegraphics[scale=0.75]{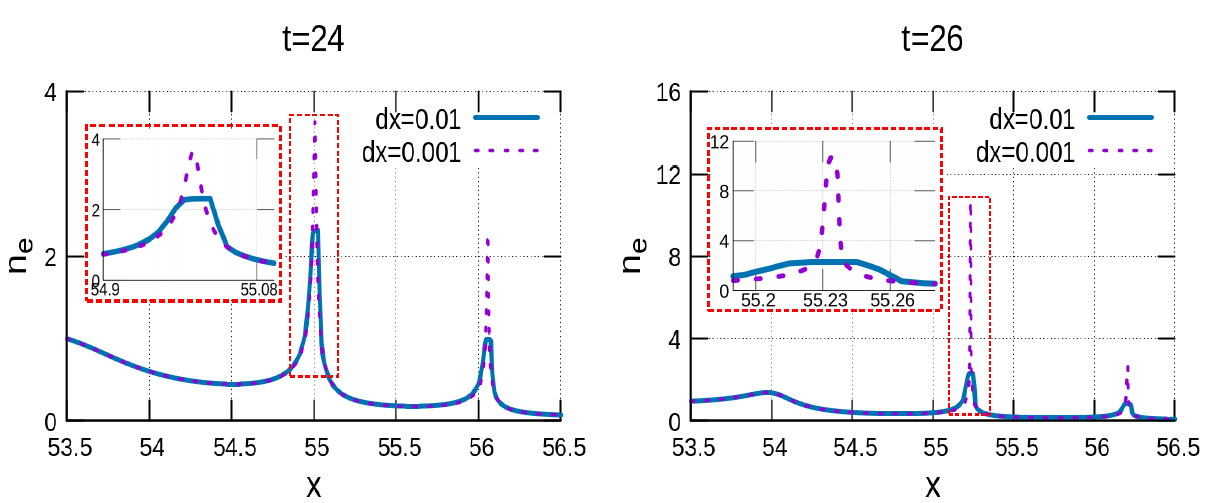}
    \caption{Density of electron corresponding t=24 and 26 for smooth profile case with dx= 0.01 and 0.001.}
    \label{fig:8}
\end{figure}

\begin{figure}
    \centering
    \includegraphics[scale=0.75]{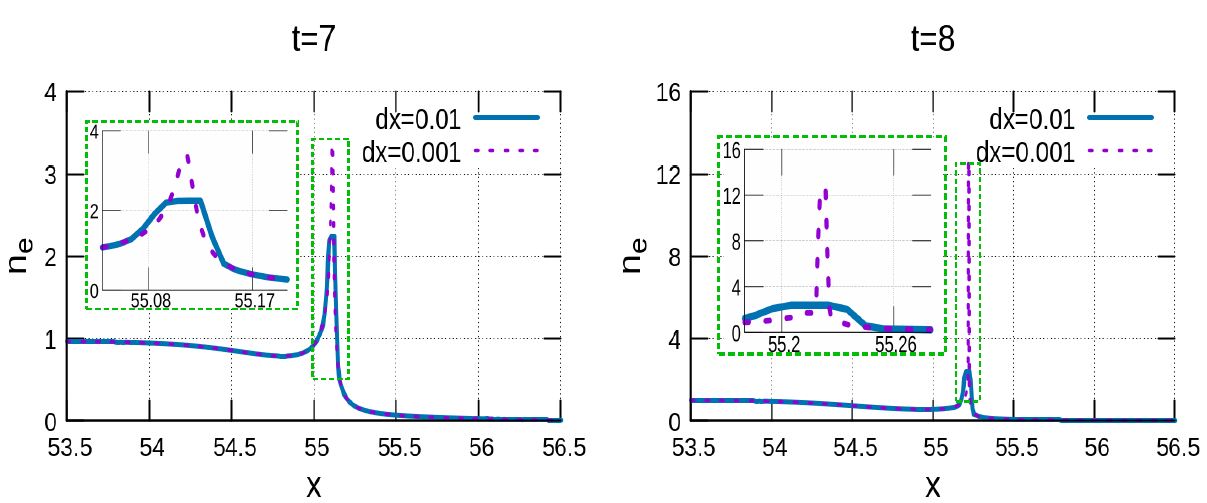}
    \caption{Density of electron corresponding t=7 and 8 for sharp profile case with dx= 0.01 and 0.001.}
    \label{fig:9}
\end{figure}

We have also plotted the structures that develop in electron density for the two choices of grid size. In figure (\ref{fig:8}), we show a comparison for the cases of two grid sizes at a time when the energy conservation has just started to fail for the cruder grid spacing 
at $t = 24$. It can be observed that the structure is identical everywhere except around the peak, where it becomes blunt at the grid scale compared to the finer grid spacing for which the energy conservation remains valid at this time. At a later time of  $t = 26$, while the energy conservation remains valid for the finer grid, a comparison has been provided in the right-hand side subplot of the figure (\ref{fig:8}).
The cruder grid size shows further broadening. 
These observations are confirmed for the choice of a sharper background plasma profile also (see figure (\ref{fig:9})).

\begin{figure}[!hbt]
     %\centering
     \begin{subfigure}[b]{0.5\textwidth}
         \centering
         \includegraphics[scale=0.34]{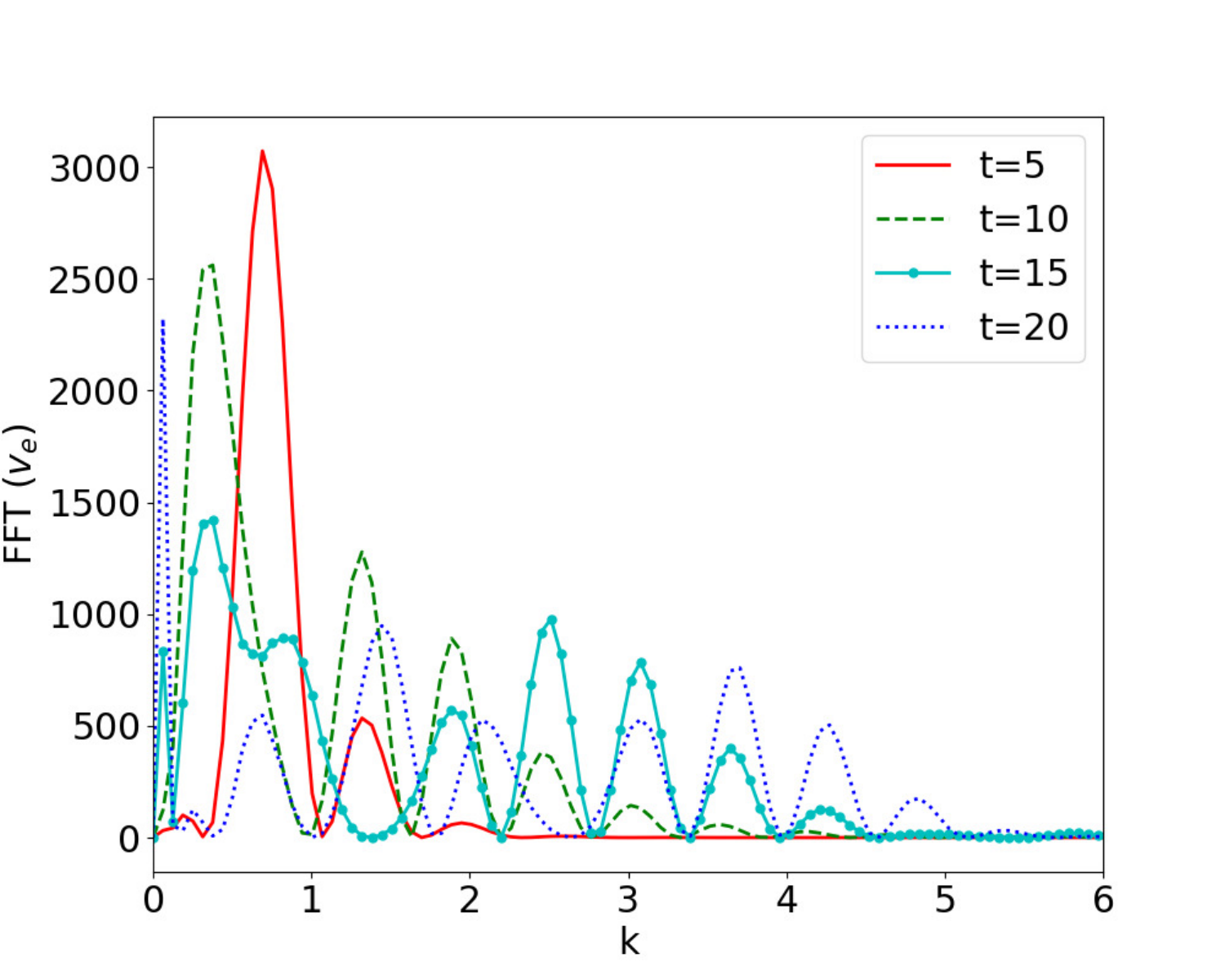}
         \caption{Spacial FFT for the smooth case at different times.}
         \label{fig:10a}
     \end{subfigure}
     %\hfill
          \begin{subfigure}[b]{0.5\textwidth}
         \centering
         \includegraphics[scale=0.34]{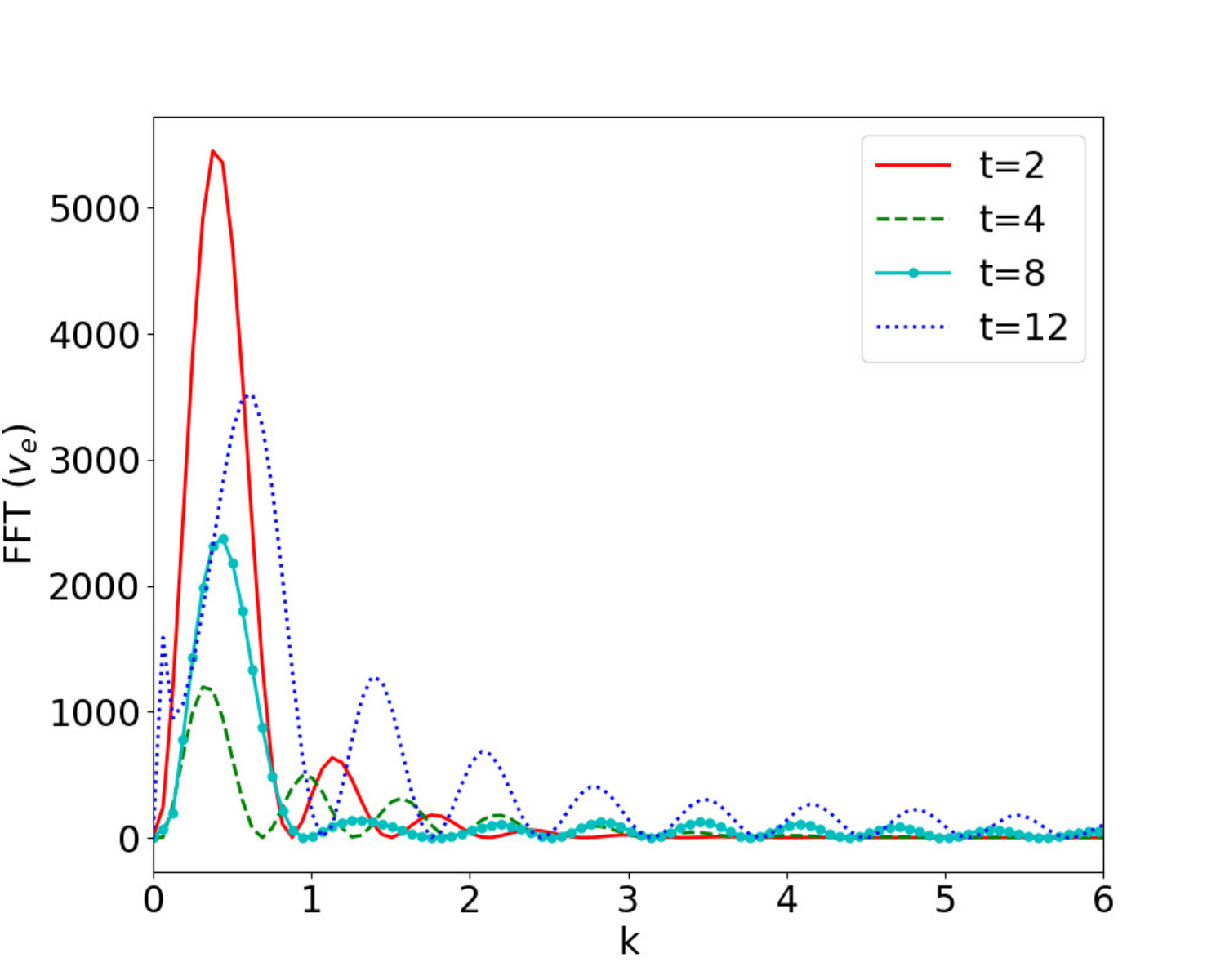}
         \caption{Spacial FFT for the sharp case at different times.}
         \label{fig:10b}
     \end{subfigure}
        
        \caption{Spatial FFT for the smooth and sharp cases.}
        \label{fig:10}
\end{figure}

\begin{figure}[!hbt]
     %\centering
        \begin{subfigure}[b]{0.5\textwidth}
         \includegraphics[scale=0.34]{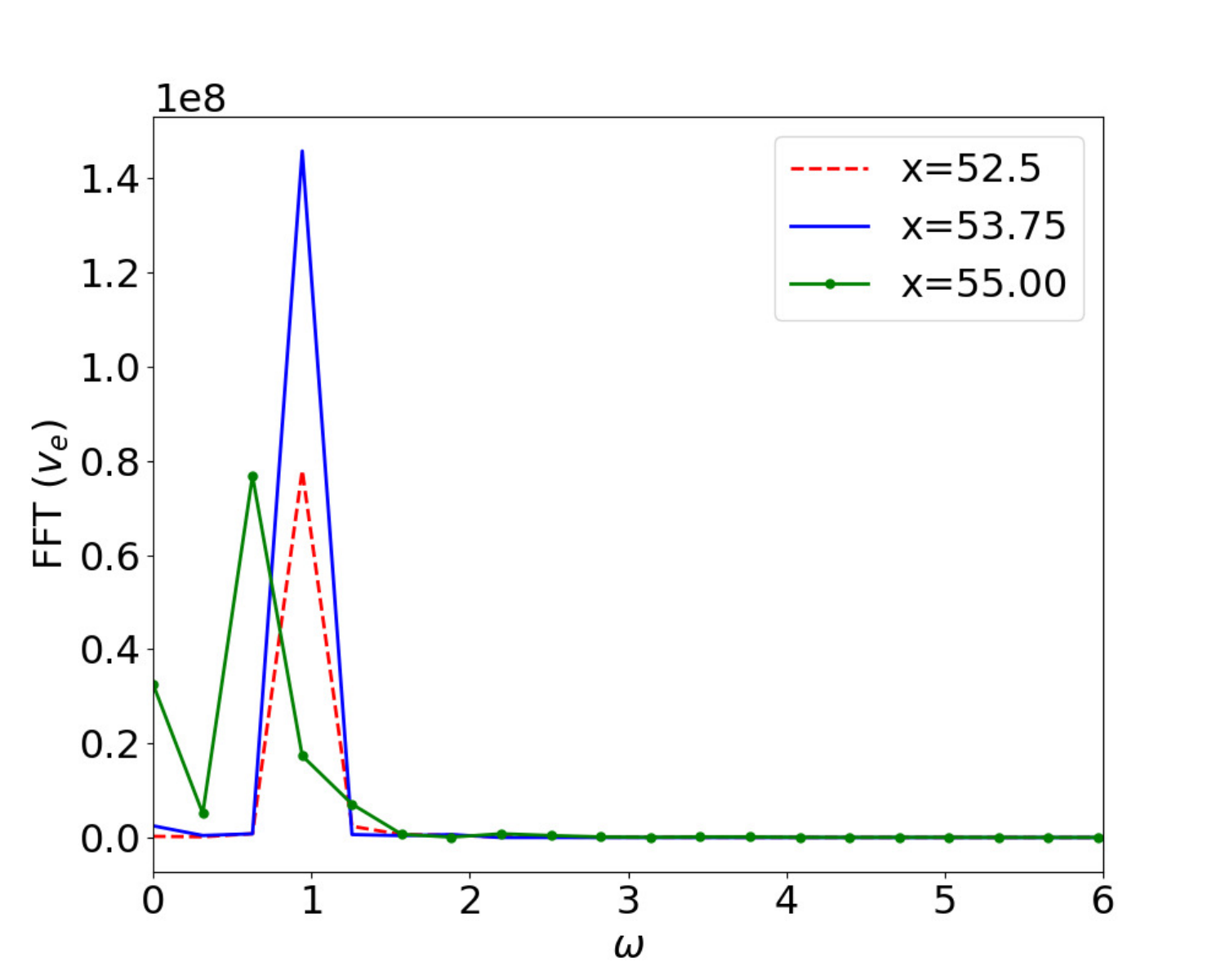}
         \caption{Temporal FFT for the smooth case at different locations.}
         \label{fig:11a}
     \end{subfigure}
        \begin{subfigure}[b]{0.5\textwidth}
         \includegraphics[scale=0.34]{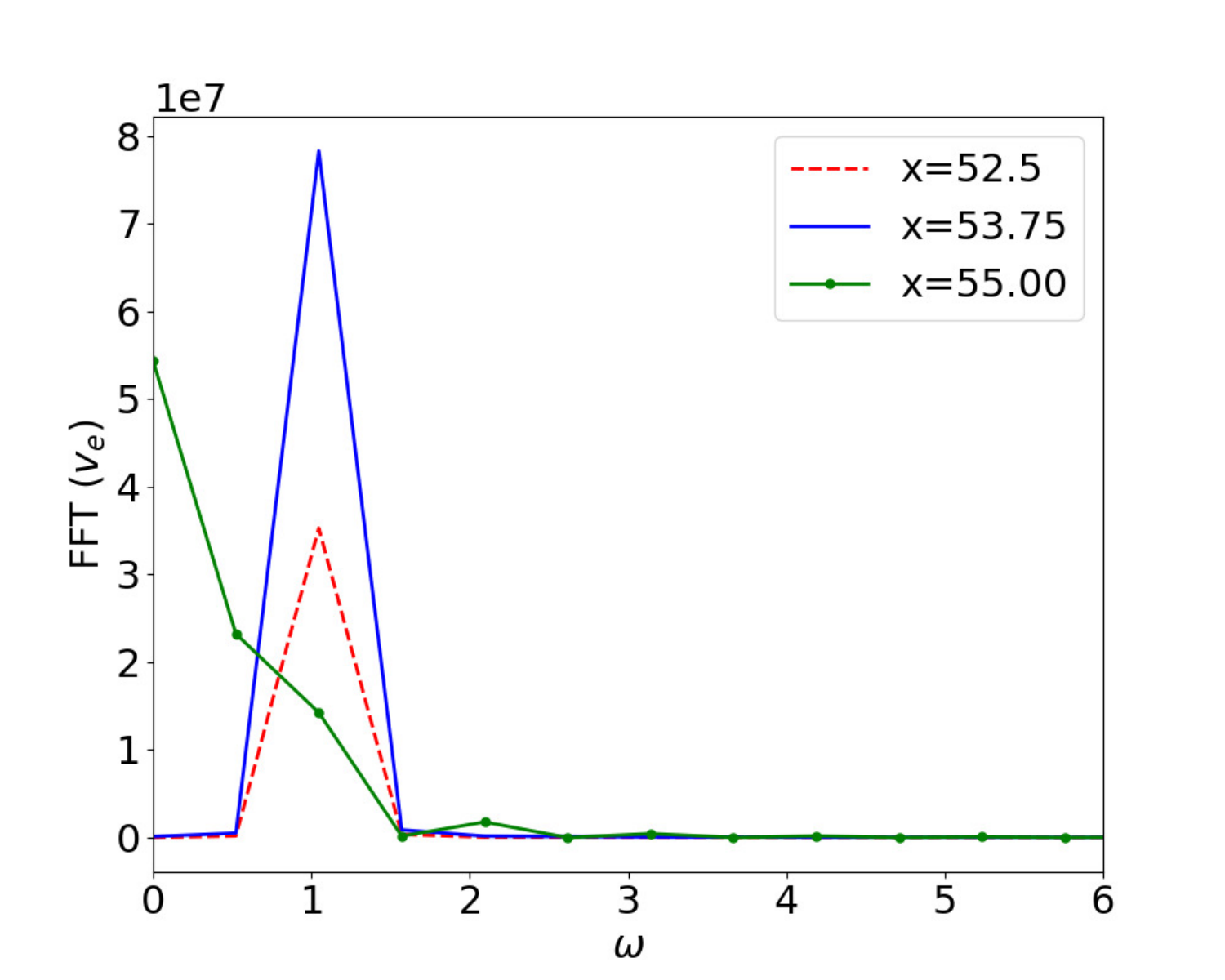}
         \caption{Temporal FFT for the sharp case at different locations.}
         \label{fig:11b}
     \end{subfigure}
        \caption{Temporal FFT for the smooth and sharp cases.}
        \label{fig:11}
\end{figure}

The tendency to form sharper structures can also be observed in density as well as other field variables. The spatial Fourier transformation of the electron velocity field has been depicted in figures (\ref{fig:10a}) and (\ref{fig:10b}). It can be observed that power in high wavenumber gets enhanced. 
Similarly, the Fourier transform in time for perturbed fields at three different spatial locations has been shown in figure (\ref{fig:11}). The frequency spectrum shows maximum power in the local plasma oscillation frequency and its harmonics.

It is worth mentioning that with the development of higher wavenumber in the perturbed fields, the energy cascades from longer to shorter scales.  The generation of higher wavenumbers and higher harmonics in the context of inhomogeneous and/or finite ion mass leads to the phenomena of phase mixing leading to wave-breaking \cite{sengupta2009phase,sengupta2011phase,bera2016relativistic}. 

It should be noted here that the improved grid resolution does not arrest the growth of sharper structures. No matter how much the grid resolution is improved upon the ultimate destiny is the formation of sharp structures wherein the power keeps going to the shortest grid scale. It should be noted that though the grid resolution was improved by $10$ fold, the additional duration for which the fluid simulation remains valid was only a couple of plasma periods. Thus the onset of failure of energy conservation in the fluid simulation here in fact is an indicator of typical phase mixing time. 
This has been adopted as an indicator of phase mixing time in some earlier publications. The phase mixing time estimate thus obtained from the fluid simulation was found to have good agreement with the analytic derivation of the phase mixing time \cite{sengupta2009phase,sengupta2011phase,bera2016relativistic}.

\begin{figure}[!hbt]
    \centering
    \includegraphics[scale=0.4]{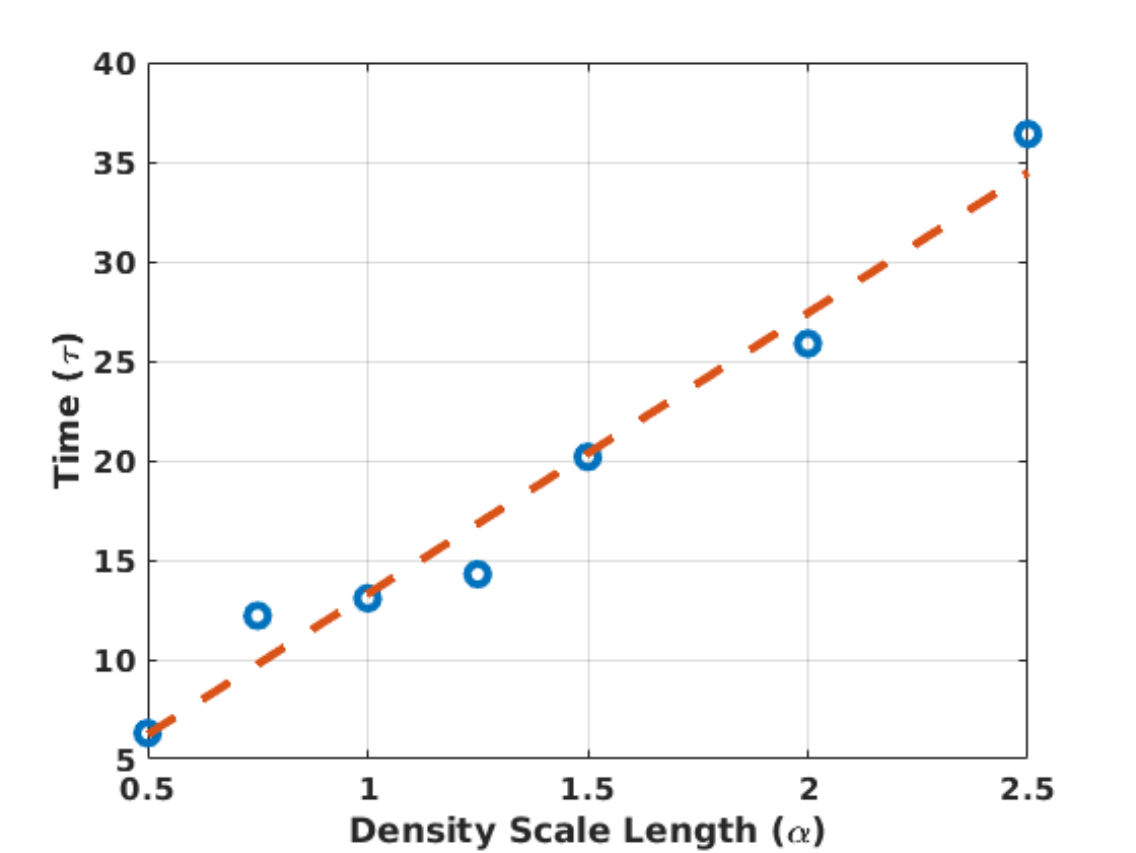}
    \caption{Time of Energy Damping ($\tau$) with density scale length at the profile edge.}
    \label{fig:12}
\end{figure}

We illustrate the dependence of the phase mixing time ($\tau$) using the breakdown of energy conservation for various sharpness determined by the parameter $\alpha$ of the background plasma profile. We note that there appears to be a linear dependence of phase mixing time ($\tau$) with $\alpha$ as shown in figure (\ref{fig:12}).
Thus, the sharpness of a finite-size plasma leads to faster wave breaking. 

The formation of sharp structures can get arrested in the presence of  
natural dissipative processes. In the next section, we introduce the effect of viscosity in our fluid simulations and show that these sharp structures now tend to settle down at a particular scale.

\section{Effect of Additional Diffusion on Plasma Evolution}{\label{4s}}

\begin{figure}[!hbt]
    \centering
    \includegraphics[scale=0.8]{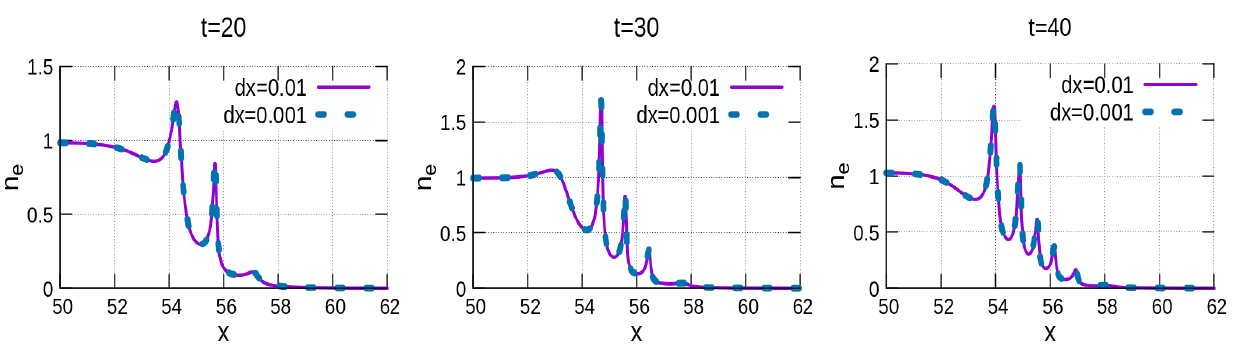}
    \caption{Electron density Evolution for $\eta=0.001$.}
    \label{fig:13}
\end{figure}

\begin{figure}[!hbt]
    \centering
    \includegraphics[scale=0.8]{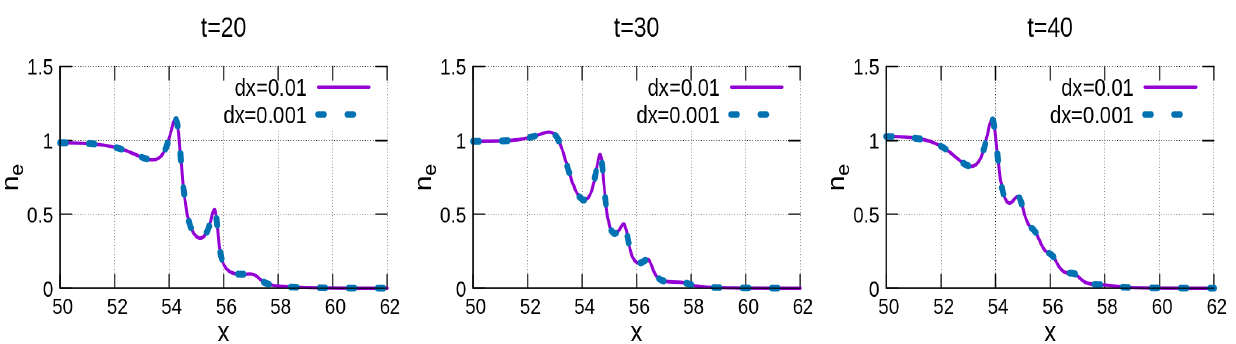}
    \caption{Electron density Evolution for $\eta=0.005$.}
    \label{fig:14}
\end{figure}

\begin{figure}[!hbt]
    \centering
    \includegraphics[scale=0.8]{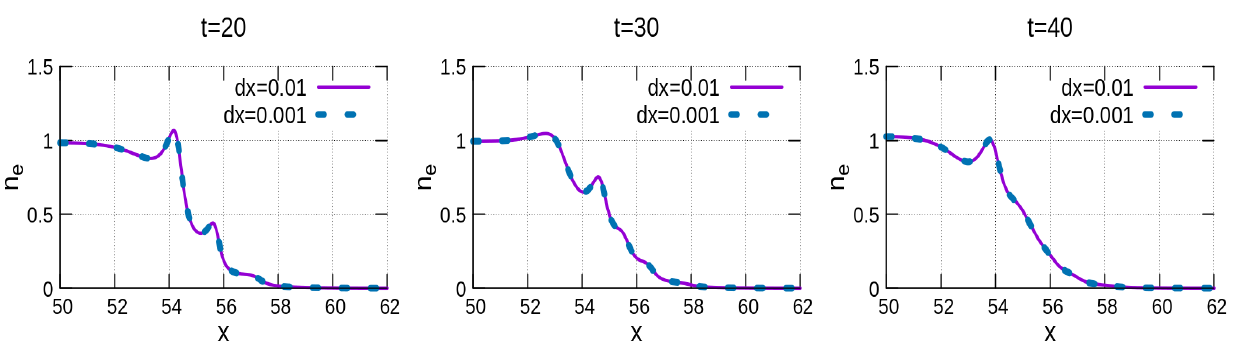}
    \caption{Electron density Evolution for $\eta=0.01$.}
    \label{fig:15}
\end{figure}

\begin{figure}[!hbt]
    \centering
    \includegraphics[scale=0.6]{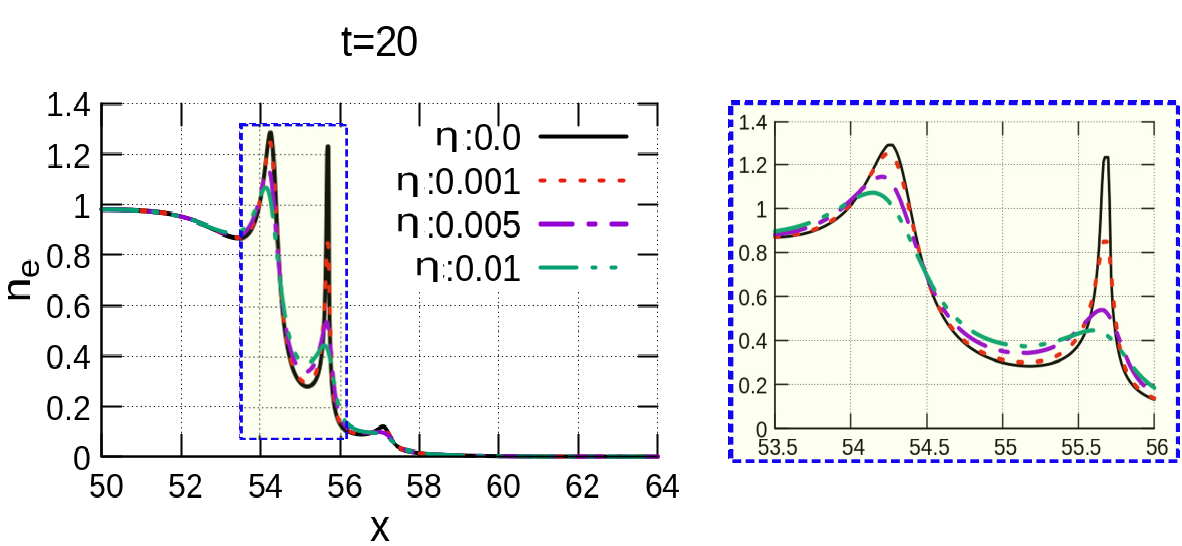}
    \caption{Electron density Evolution for different $\eta$ at t=20.}
    \label{fig:16}
\end{figure}

\begin{figure}[!hbt]
    \centering
    \includegraphics[scale=0.6]{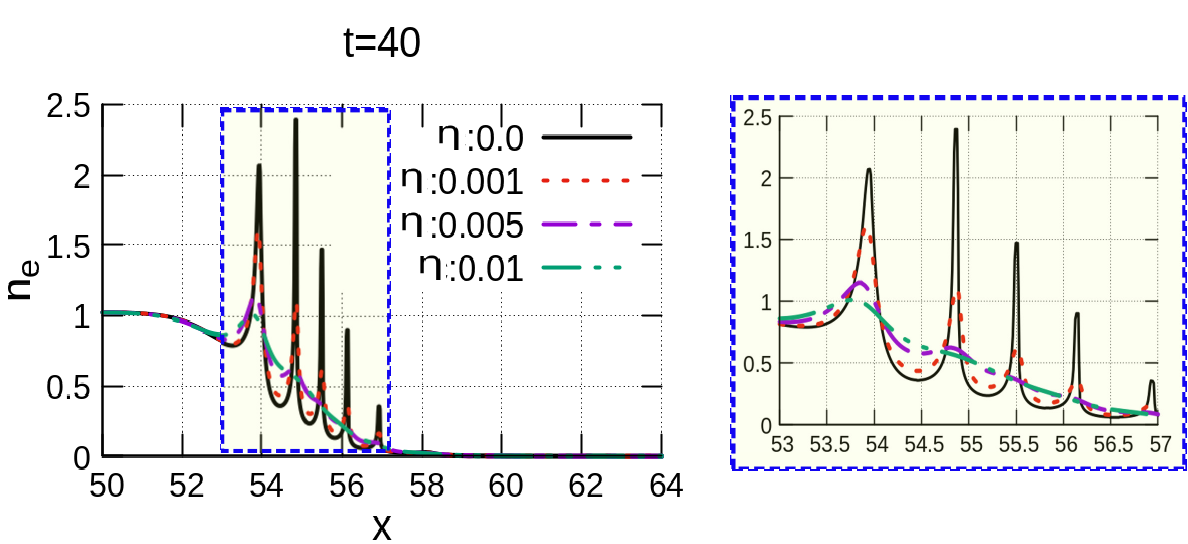}
    \caption{Electron density Evolution for different $\eta$ at t=40.}
    \label{fig:17}
\end{figure}

In this section, we will demonstrate that the addition of dissipation arrests the sharpening of these structures. 
We modify the electron momentum equation and include  viscous dissipation to have Eq.(\ref{eq:8}) below:
\begin{equation}\label{eq:15}
    {\frac{\partial{u}_e}{\partial{t}}}+{u}_e{\frac{\partial{u}_e}{\partial{x}}}={\frac{\partial}{\partial{x}}}\left({\phi}+{\eta}\frac{\partial{u}_e}{\partial{x}}\right)
\end{equation}
Here $\eta$ is the coefficient of viscosity. We now show that for an appropriate choice 
of the viscosity parameter, the sharpness of the structures can be arrested at scales longer than the grid scale, and refining the grid does not lead to any difference. In figures (\ref{fig:13}), (\ref{fig:14}) and (\ref{fig:15}) we show the plot for two different grid sizes ($dx = 0.01$ and $0.001$) at various times. The structures for different grid resolutions remain identical. This demonstrates that the presence of natural dissipative processes  
does not allow the structures to develop sharpness beyond a certain point. The fluid depiction thus continues to remain valid in such a case. 

We also provide a comparison of the plasma density profile for various values of the viscosity parameter in figures (\ref{fig:16}) and (\ref{fig:17}). It can be seen that as expected a higher value of $\eta$ leads to the formation of a broader structure. 

The natural dissipative processes  lead to energy dissipation and the decay of total energy is given  by the following equation:
\begin{equation}\label{eq:16}
    {\frac{d}{d{t}}\int \left[\frac{1}{2}{n}_e{u}_e^2+\frac{1}{2}\frac{m_i}{m_e}{n}_i{u}_i^2+\frac{1}{2}\left(\frac{\partial{\phi}}{\partial{x}}\right)^2 \right]d{x}= \frac{d}{dt} E_t = -{\eta}\int\left(\frac{\partial{u_e}}{\partial{x}}\right)^2d{x}}
\end{equation}

\begin{figure}[!hbt]
    \centering
    \includegraphics[scale=0.75]{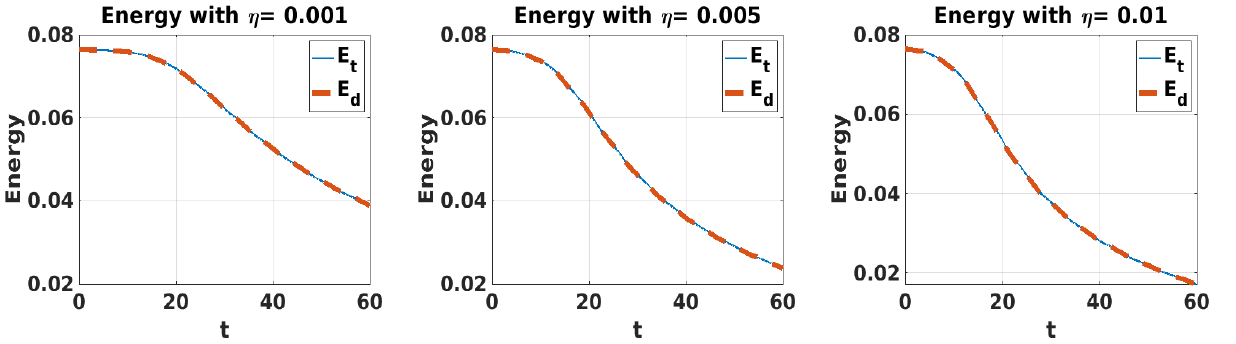}
    \caption{Total energy with viscous effect for different $\eta$.}
    \label{fig:18}
\end{figure}

We have shown the energy evolution from simulation by using the expression of total energy $E_t$. This total energy decreases right from the very beginning due to the presence of the dissipative term on the right-hand side. The total energy evolution is compared with a plot of the energy decay arising from the right-hand side of the equation in figure (\ref{fig:18}). We can observe that the two plots are the same.

We now focus on understanding the process of expansion of any of the finite-sized plasma. We measure the expansion rate by evaluating the Mean Square Displacement (MSD) as a function of time for both electron and ion fluids. It is  defined by  (\ref{eq:17})
\begin{equation}\label{eq:17}
     \langle X_s^2 \rangle =\frac{\int(x-x_0)^2 {n_s} dx}{\int{n_s}dx}
\end{equation}
Here the suffix $s = e, i$ stands for electron and ion fluids, respectively, and $x_0$ is the center of their background density profile. 
The time evolution of $\langle X_s^2\rangle $ thus characterizes a typical expansion of the density profile of the species $s$.

\begin{figure}[!hbt]
     \begin{subfigure}[b]{0.5\textwidth}
         \includegraphics[scale=0.4]{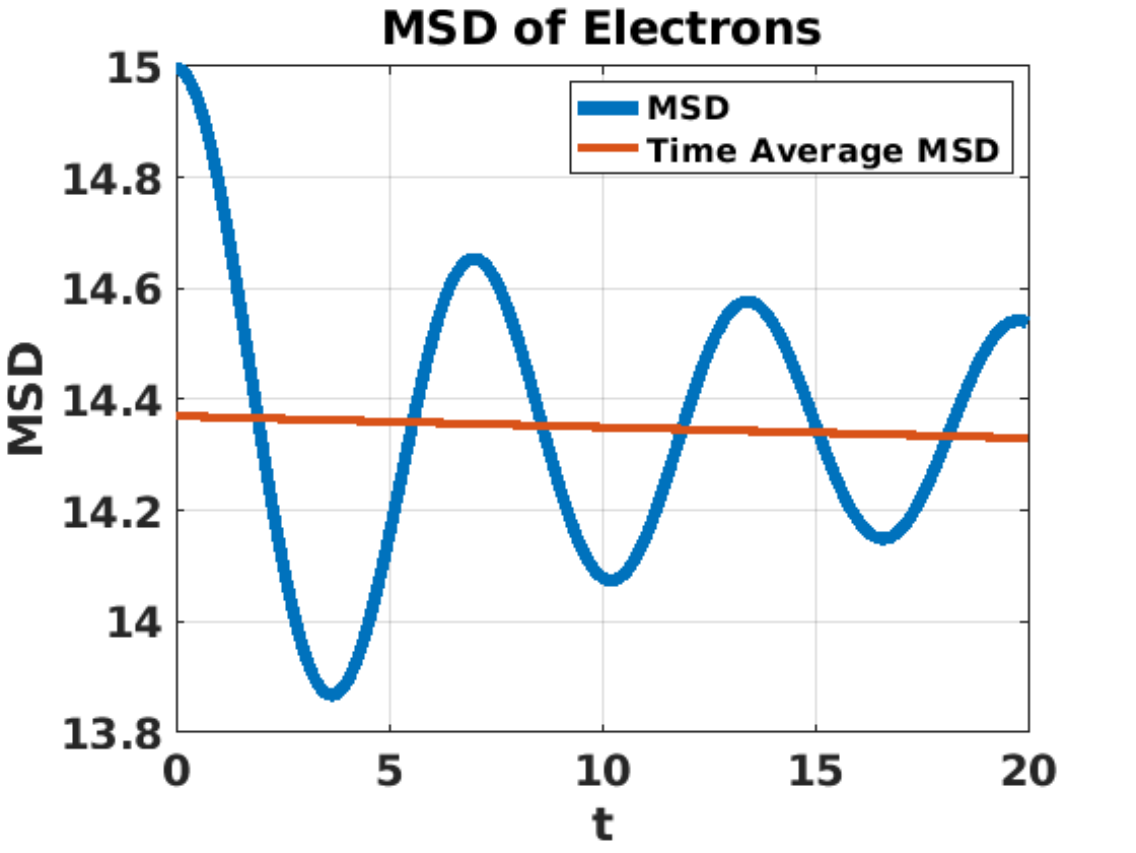}
         \caption{}
         \label{fig:19a}
     \end{subfigure}
     \begin{subfigure}[b]{0.5\textwidth}
         \includegraphics[scale=0.415]{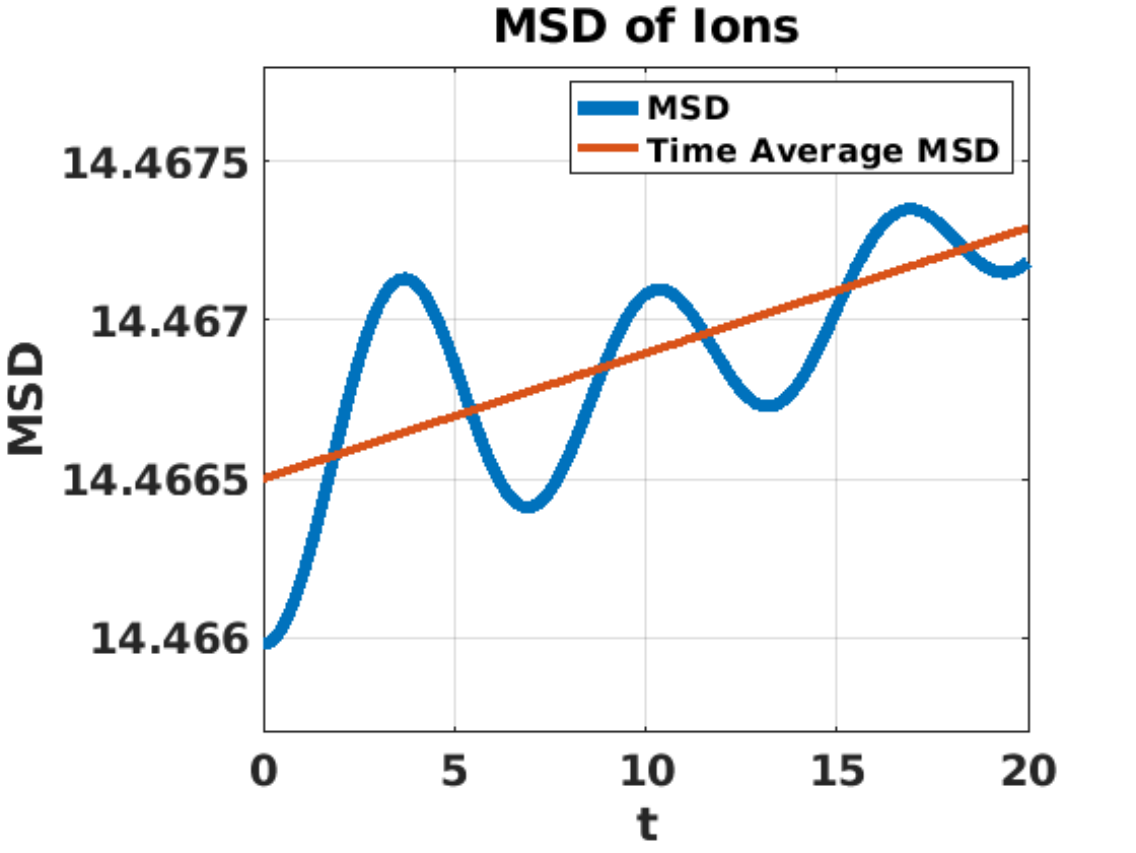}
         \caption{}
         \label{fig:19b}
     \end{subfigure}
     \caption{Mean Square Displacement (MSD) of electron and ion fluid.}
    \label{fig:19}
\end{figure}

\begin{figure}[!hbt]
     \begin{subfigure}[b]{0.5\textwidth}
         \includegraphics[scale=0.4]{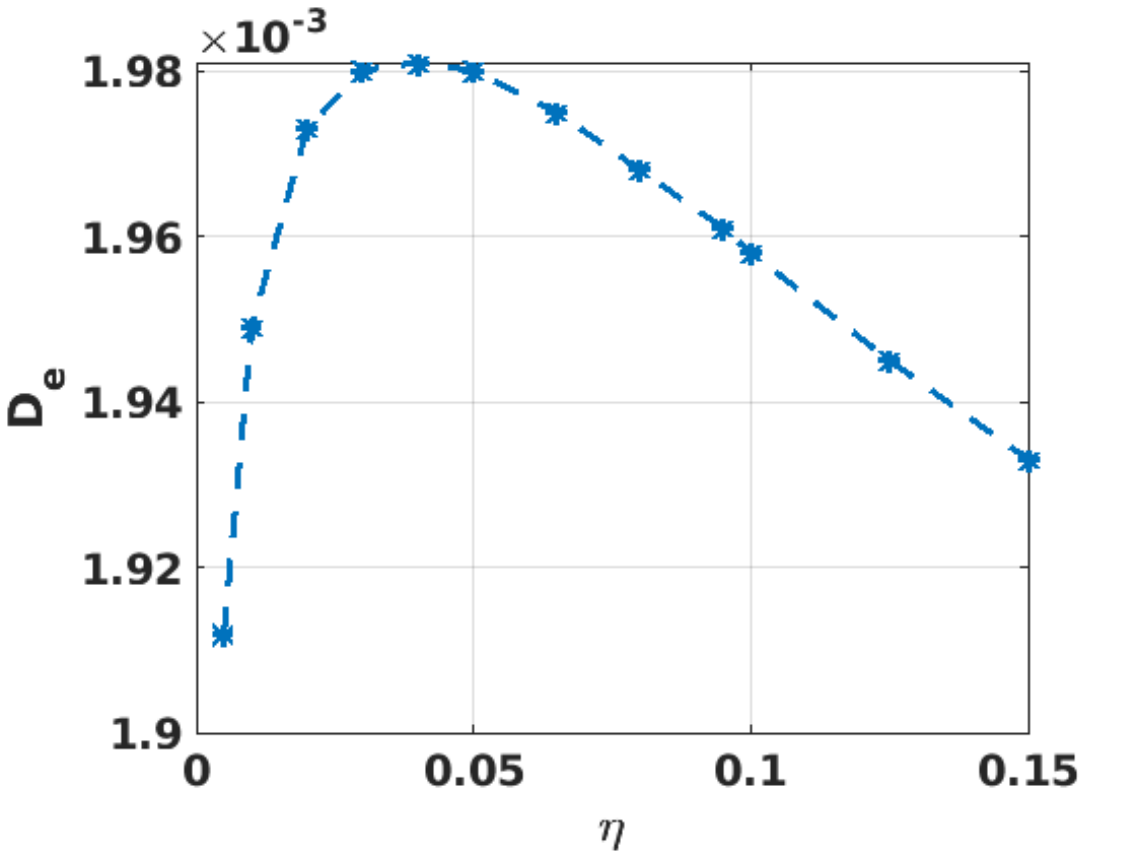}
         \caption{}
         \label{fig:20a}
     \end{subfigure}
     \begin{subfigure}[b]{0.5\textwidth}
         \includegraphics[scale=0.4]{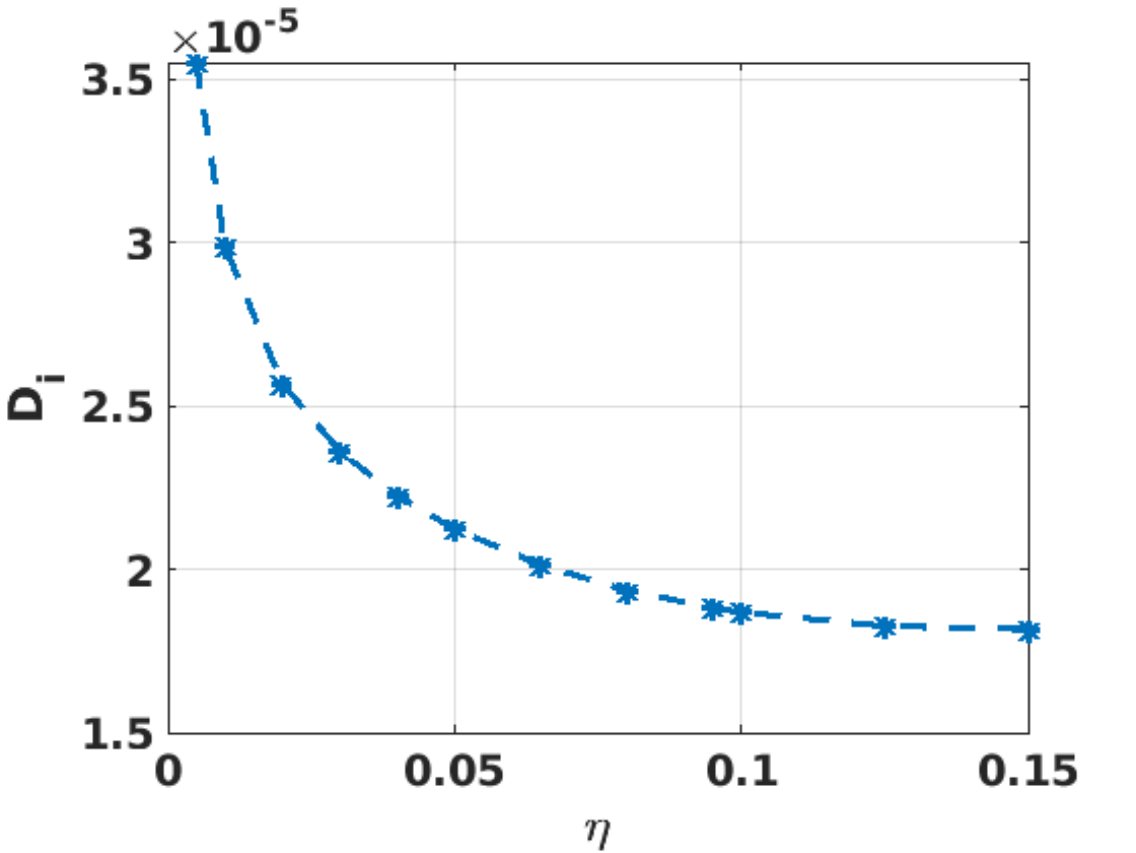}
         \caption{}
         \label{fig:20b}
     \end{subfigure}
     \caption{Mean Square Displacement D of electron ($D_e$) and ion ($D_i$) fluid with the external diffusion constant $\eta$.}
    \label{fig:20}
\end{figure}

Figures (\ref{fig:19a}) and (\ref{fig:19b})  show the evolution of electron and ion MSD for the case when $\eta=0$ for  $\alpha = 1.5$. 
The presence of out-of-phase oscillations in $\langle X_s^2\rangle $ for electrons and ions is evident from the plots. However, upon time averaging, it is clear that the electrons that had a broader profile compared to ions initially tend to shrink as there is a steady decrease in $ \langle X_e^2\rangle $. For ions, however,  there is a slow expansion. 

An effective diffusion coefficient $D_s$ can be obtained from the slope of the averaged MSD plotted concerning time which is a linear curve.

We have evaluated $D_s$ even for a finite value of $\eta$. The dependence of 
$D_s$ on $\eta$ has been shown in figure (\ref{fig:20}) From the plot, it is clear that by increasing the value of $\eta$ the transport coefficient of electrons first increases and then decreases. On the other hand for ions, the value keeps reducing with  $\eta$. The decrease of transport at higher values of $\eta $  can be understood by realizing that with increasing $\eta$ the perturbed electron velocity gets significantly damped and is unable to expand the plasma. On the other hand in the other regime of low $\eta$ values the damping of the kinetic energy remains insignificant but the phase relationship between the velocity and density fields gets altered from that of plasma oscillations which have a reversible characteristic. 
This introduces additional randomness and is responsible for increasing the rate of plasma expansion.

In the next section, we study the role of external forcing on the evolutionary behavior of finite-size plasma droplets.  

\section{Effect of External Forcing}{\label{5s}}
For simplicity, the forcing is chosen to have a form of dipole electric field. This can also be looked upon as a long-wavelength incident radiation field. Thus we choose the form of the electric field as $E=E_0cos(\omega_{L}{t})$. Here $E_0$ is the amplitude of the external applied field having a frequency $\omega_L$.  The initial profile of plasma density for both electrons and ions is chosen to be identical for these driven simulations. Thus,  there is no self-consistent electric field to begin with. Plasma dynamics is governed entirely by the external electric field in this case. 

\begin{figure}[!hbt]
    \centering
    \includegraphics[scale=0.25]{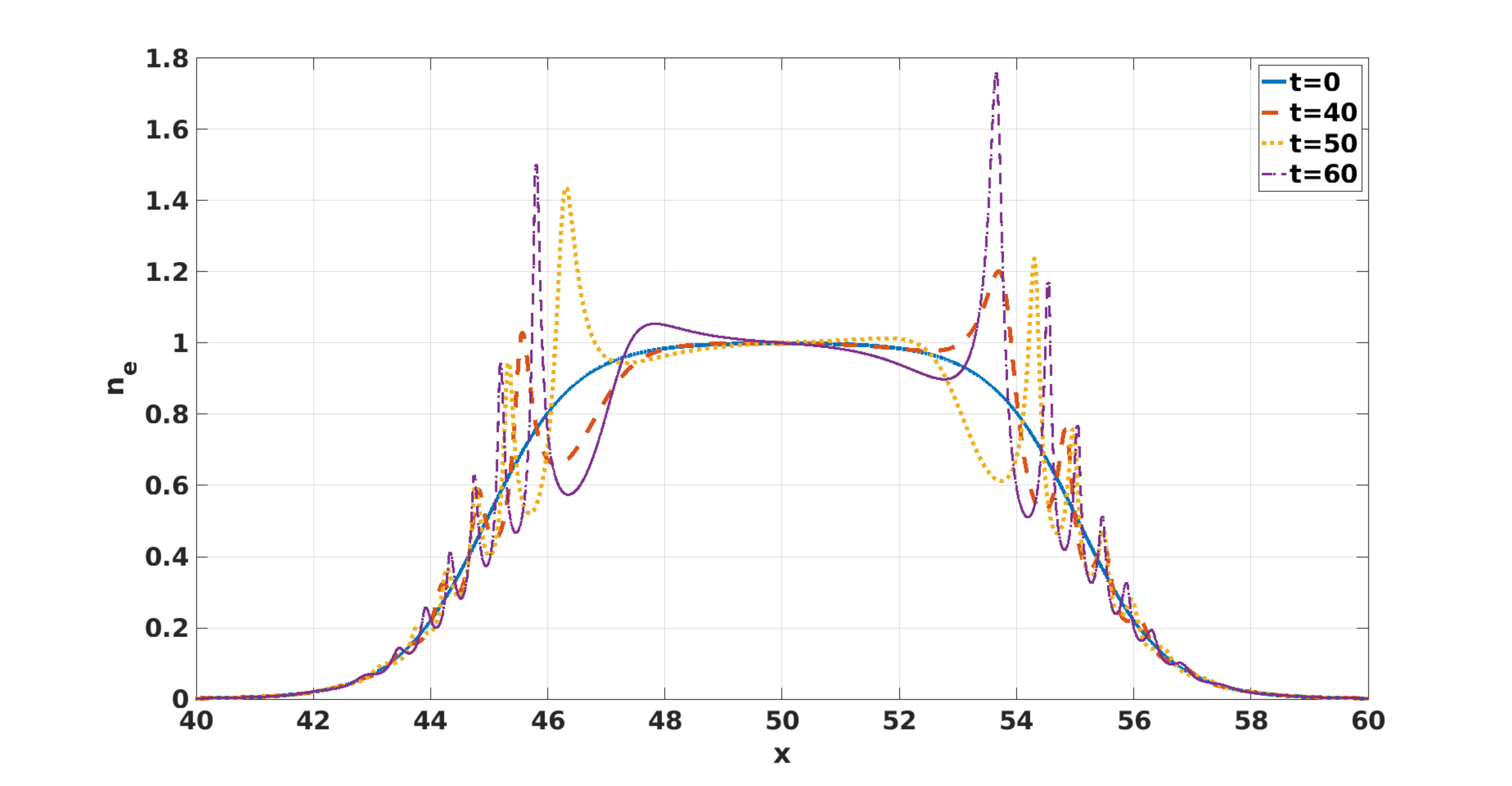}
    \caption{Electron density evolution for the external electric field $E=0.01cos(t)$.}
    \label{fig:21}
\end{figure}

\begin{figure}[!hbt]
    \centering
    \includegraphics[scale=0.25]{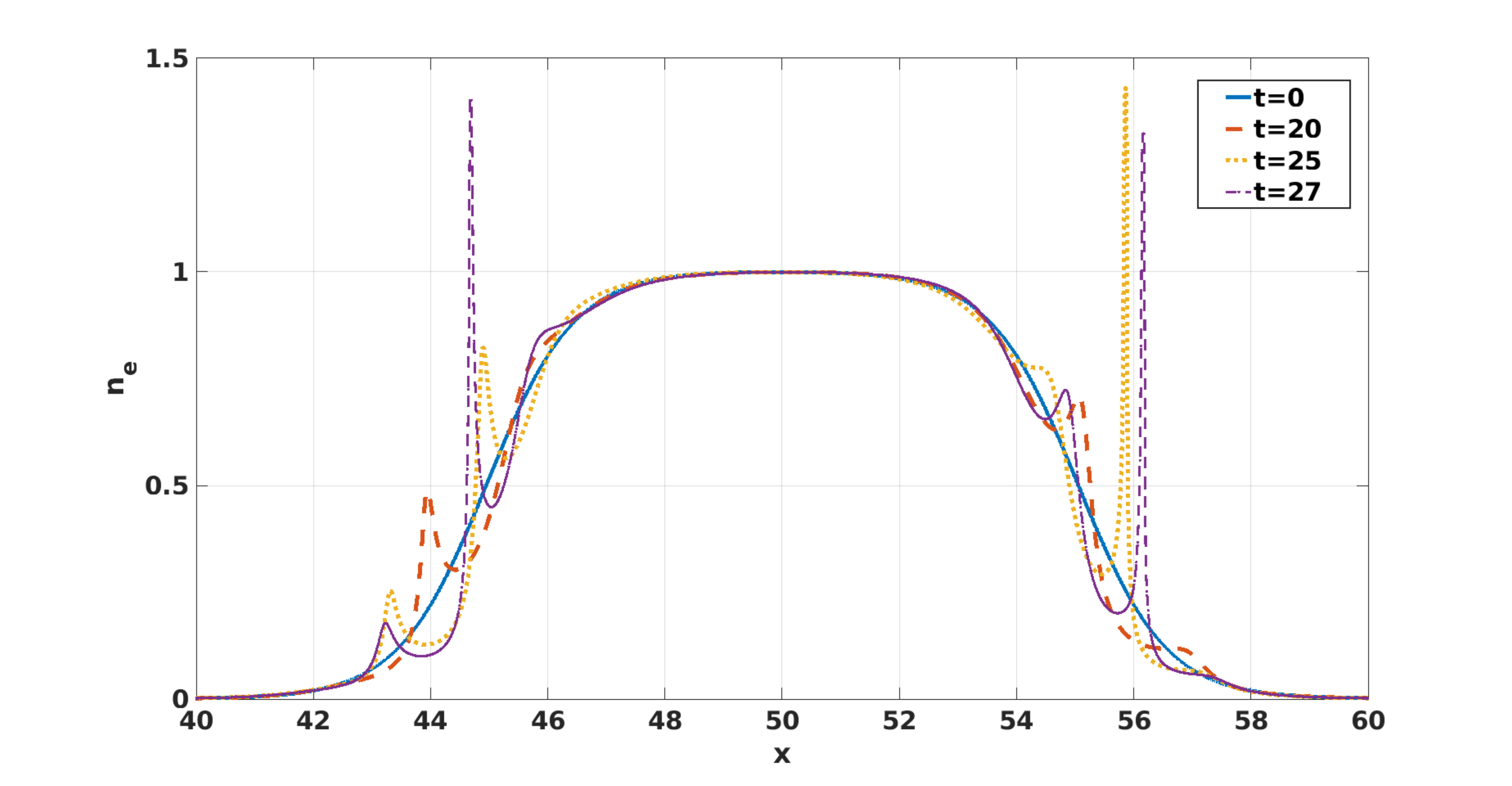}
    \caption{Electron density evolution for the external electric field $E=0.01cos(0.5t)$.}
    \label{fig:22}
\end{figure}

\begin{figure}[!hbt]
    \centering
    \includegraphics[scale=0.25]{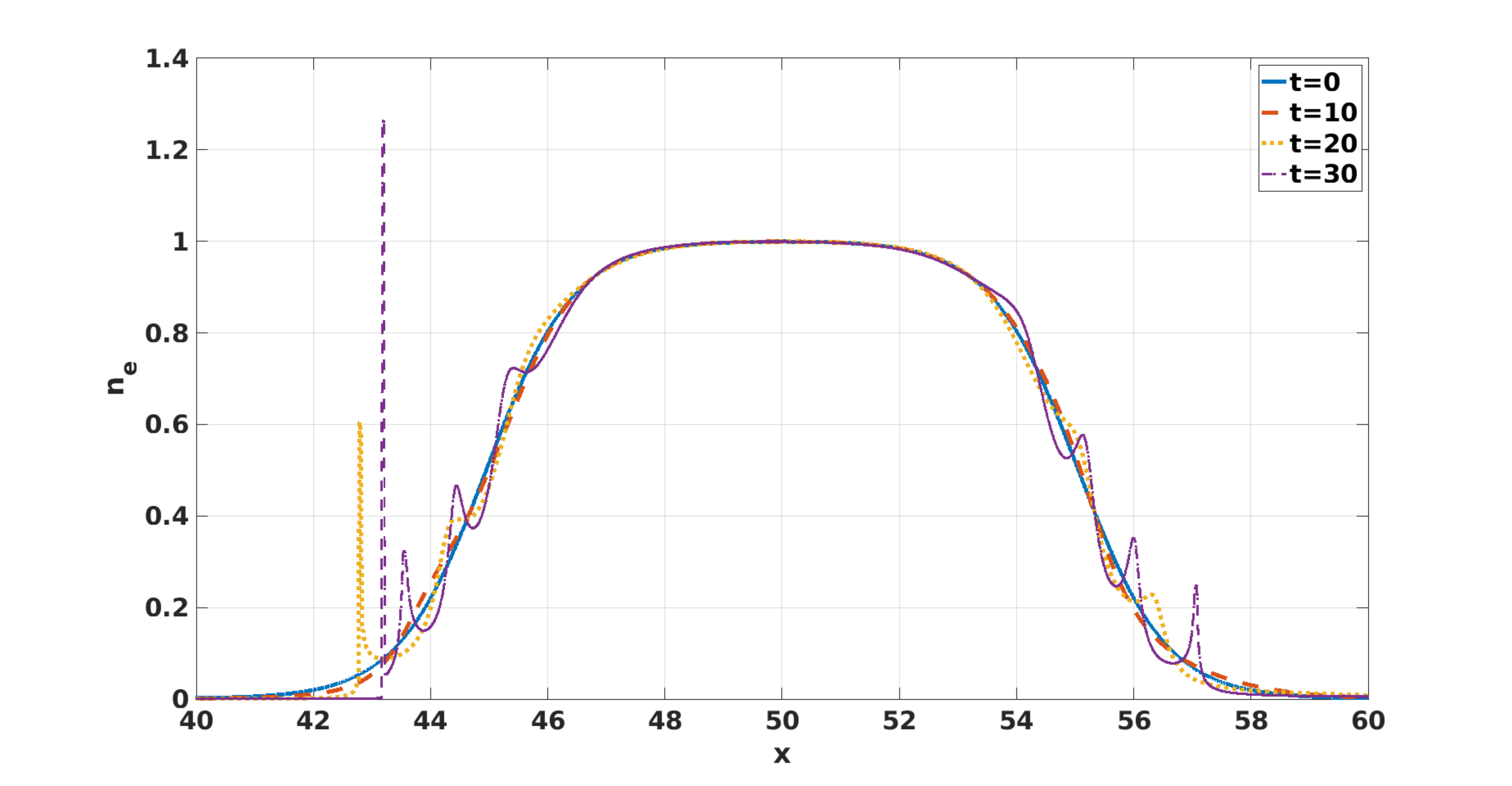}
    \caption{Electron density evolution for the external electric field $E=0.01cos(0.01t)$.}
    \label{fig:23}
\end{figure}

We first choose a small amplitude of the driving field for which $E_0 = 0.01$. We consider $\omega_L = 1$ and $\omega_L = 0.5$. The evolution of the density profile is shown as snapshots in figures (\ref{fig:21}) and (\ref{fig:22}), respectively for these cases. In the supplementary material, we have attached the movie for these cases. There is an asymmetric development of the electron density perturbations. The ions at this frequency remain unperturbed. 
For  $\omega_L = 1$,  the external driver resonates with the maximum density at the flat top central region.  The electron density perturbation here originates just at the edge of the flat top region of the density profile. In each cycle, the perturbations appear alternately on both sides. Subsequently, they move down the density slope.  For the lower frequency of $\omega_L = 0.5$ one observes that the perturbations appear again at the resonant point which in this case is located at the edge. We have also studied cases with several lower values of the driver frequency and observed that the perturbations are essentially initiated at the local resonance point. These perturbations get sharper with time and also move down the gradient in all cases. Thus in the presence of an external driver, both the phenomena of resonance and the movement of perturbations towards shallower density are operative. 

\begin{figure}[!hbt]
    \centering
    \includegraphics[scale=0.8]{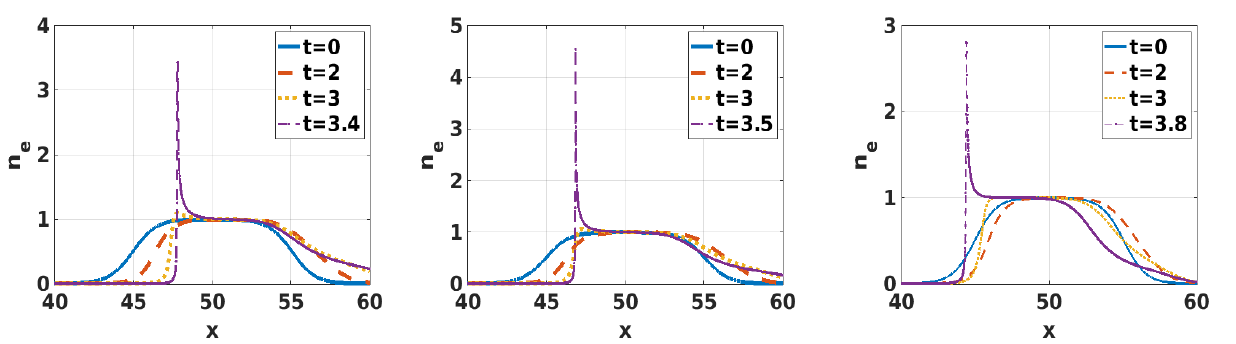}
    \caption{Electron density evolution for the external electric field $E=cos(0.01t)$, $E=cos(0.5t)$, $E=cos(t)$.}
    \label{fig:24}
\end{figure}

At very low frequencies, the electrons essentially see a DC field. In this case, the dominant perturbations continue to remain only at one side of the profile where the electrons are pushed out. This can be observed from the figure (\ref{fig:23}). When the frequency of the driver is chosen to be much larger than the plasma frequency, there is no resonance. Also, even the lighter electron species are unable to respond. This case does not yield anything interesting. 
Increasing the amplitude of the driver field results in a faster appearance of large amplitude perturbations (figure (\ref{fig:24})). 

\section{Summary} {\label{6s}}
We have carried out 1-D fluid simulations to study the evolution of the finite-size plasma droplet. Unlike extended plasmas in this case no matter how small the initial charge density perturbation one chooses it results in the formation of sharp structures that form at the edge region and keep becoming sharper as they move down the density gradient. The central flat-top region, however, remains unperturbed throughout. The dynamics and wave breaking remain confined in the edge inhomogeneous density. 

For a driven system, the eternal field itself generates density perturbations leading to charge imbalance and a self-consistent field. In this case, however, the perturbations get initiated at the location where the external driver frequency resonates with the local plasma frequency. The disturbances here too move down the density gradient and keep getting sharper. Thus in the presence of an external driver, both resonance and movement of perturbations down the gradient play a role.

\bibliographystyle{unsrt}
%\bibliography{library}

\end{document}